\documentclass[11pt]{iopart}

\usepackage{iopams}
\usepackage{amsfonts,bbm,amssymb}
\usepackage{setstack}
\usepackage{dsfont}
\usepackage{hyperref}
\usepackage[toc,page]{appendix}

\def\bra#1{\mathinner{\langle{#1}|}}
\def\ket#1{\mathinner{|{#1}\rangle}}
\def\expect#1{\langle#1\rangle}
\def\ol#1{\overline{#1}}
\def\bb#1{\mathbf{#1}}
\def\mbb#1{\mathbb{#1}}
\def\adH{\,{\rm \hat{ad}}_{H}}
\def\Uqsl#1{\mathcal{U}_{q}(\mathfrak{sl}(#1))}

\newcommand{\braket}[2]{\langle #1 \vert #2 \rangle}

\newcommand{\ii}{ {\rm i} }

\newcommand{\ZZ}{\mathbb{Z}}

\newcommand{\CC}{\mathbb{C}}

\newcommand{\half}{\frac{1}{2}}
\newcommand{\LL}{{\hat{\cal L}}}
\newcommand{\DD}{{\hat{\cal D}}}
\newcommand{\VV}{{\hat {\cal V}}}
\newcommand{\GG}{{\cal G}}

\def\Re{{\,{\rm Re}\,}}

\def\PR{\check{R}}
\def\sinc#1{\,{\rm sinc}(#1)}

\eqnobysec

\usepackage{fancyhdr}
\begin{document}

\title[Quantum group approach to steady states]
{Quantum group approach to steady states of boundary-driven open quantum systems}
\author{Enej Ilievski}

\address{Department of Physics, Faculty of Mathematics and Physics, University of Ljubljana,
Jadranska 19, SI-1000 Ljubljana, Slovenia}
%\ead{enej.ilievski@fmf.uni-lj.si}

\author{Bojan {\v Z}unkovi{\v c}}

\address{Departamento de F\'{\i}sica, Facultad de Ciencias F\'{\i}sicas y Matem\'{a}ticas,
Universidad de Chile, Casilla 487-3, Santiago, Chile
}


\date{\today}

\begin{abstract}
We present a systematic approach for constructing steady state density operators of Markovian dissipative evolution for open quantum chain models
with integrable bulk interaction and boundary incoherent driving. The construction is based on fundamental solutions of the quantum Yang-Baxter
equation pertaining to quantum algebra symmetries and their quantizations ($q$-deformations). 
In particular, we facilitate a matrix-product state description, by resorting to generic spin-$s$ infinite-dimensional solutions associated
with non-compact spins, serving as ancillary degrees of freedom. 
%Stationary state density matrix is cast as manifestly positive hermitian Cholesky-type factorized form, defined in terms of an non-hermitian matrix-product operator which is essentially a transfer matrix of an abstract quantum system.
After formally deriving already known solutions for the anisotropic spin-$1/2$ Heisenberg chain from first symmetry principles, we
obtain a class of solutions belonging to interacting quantum gases with $SU(N)$-symmetric Hamiltonians, using a restricted set of
incoherent boundary jump processes, and point out how new non-trivial generalizations emerge from twists of quantum group structures.
Finally, we discuss possibilities of analytic calculation of observables by employing algebraic properties of associated auxiliary vertex operators.
%By employing algebraic properties of associated auxiliary vertex operators, we particularly demonstrate that expectation values of
%particle current densities relate to the ratio of nonequilibrium partition functions for auxiliary processes. % of two successive sizes, in analogy to classical exclusion processes, and conclude with discussion of some possible further directions.
\end{abstract}
\maketitle
\tableofcontents
\newpage
\section{Introduction}
\label{sec:Intro}
Intrinsic complexity of quantum configuration space, typically preventing accurate numerical calculation or
efficient classical simulation of strongly correlated many-body systems, is one of main reasons why exact analytical solutions, despite
their scarceness, will always have a mesmerizing role among theoretical physicists.
Before formulation of the theory of integrability with rigid algebraic framework we have witnessed several
remarkable solutions of bona fide many-body quantum systems, emerging with famous Bethe's coordinate ansatz for solutions of magnetic chains in
one dimension, subsequently continuing with Onsager's work on Ising model, Baxter's solvable 2D classical vertex models, %~\cite{Baxter}
and theory of factorizable scattering matrices. A significant breakthrough was initiated
in late '60s with solution of the Kortweg-de Vries equation, soon formalized by Lax and others via $L-A$ matrix pair formulation
or analogous zero-curvature representation, %~\cite{Babelon}.
which matured with a Hamiltonian formalism ($r$-matrix approach) in the early '70s giving birth to the classical theory of solitons (being called
classical inverse scattering method), predominantly developed in Leningrad's school of mathematical physics \cite{FT}, which has been able to
resolve few other paradigmatic nonlinear partial differential equations (nonlinear Schr\"{o}dinger equation, Sine-Gordon model etc.)
within unified algebraic framework. It's  'quantization' has finally led to the \textit{quantum inverse scattering method} (QISM),
often being simply called the Algebraic Bethe Ansatz (ABA)\cite{Korepin}, which is nowadays one of main pillars of the modern theoretical
physics, finding its applications in condensed matter physics, quantum information theory and quantum field theories.

Since objects from QISM are merely a special (symmetry-enhanced) form of matrix product states, the latter playing a profound role in
efficient description of quantum states in cutting edge numerical simulations, it is of no surprise that Baxter's and Faddeev's work strongly
influenced the ongoing progress in the area of exact solvability.
Few of the most prominent solutions belong in the realm of classical exclusion processes and reaction-diffusion processes in
the context of classical master equations, and e.g. valence-bond states in the quantum domain~\cite{AKLT,Klumper}.
However, in spite of all those major developments in the area of quantum integrability,
until very recently surprisingly no \textit{exact many-body} solutions of  \textit{non-equilibrium quantum} dynamics have appeared in the literature.

In the seminal work of Prosen, a general method for exact diagonalization of quadratic (quasi-free) Liouville operators in terms
of normal decay modes within Markovian quantum master equation framework, by performing operator quantization over Liouville-Fock space,
has been introduced~\cite{3Q}, and then further developed in \cite{TB,BT}. Interacting systems are much harder to deal with, and at the moment we are still lacking
exact many-body solutions of full Liouvillian spectrum for a genuinely interacting many-body system.
Therefore, it is reasonable to start addressing first the simplest and perhaps physically most interesting cases
of Liouville eigenmodes, namely the steady state density operators, i.e. time-asymptotic states (fixed points) of Liouvillian dynamics.
First solution in this direction has been presented in matrix-product formulation for a non-interacting spin chain
with non-trivial bulk dephasing noise in~\cite{Marko}.
Ultimately, Prosen devised an exact ansatz for the axially-anisotropic Heisenberg spin chain in one spatial dimension
driven away from equilibrium regime via two boundary (incoherent) jump processes acting on leftmost and rightmost
particles only~\cite{PRL106,PRL107,asymmetric}.
Steady state density operator has been cast in a \textit{matrix-product operator} (MPO) form via auxiliary hopping process,
reminiscent to previously known matrix-product solutions of asymmetric classical exclusion processes (ASEP)~\cite{Derrida,Blythe,Schutz}.
Although at first glance solutions appeared rather mystical, predominantly as they were obtained in a crude ad-hoc fashion and thus
provided only little insight into the structure of the problem, they naturally called for deeper theoretical understanding. % and demanded more concise explanation.
A significant progress in this respect has been made by Karevski \etal\cite{KPS}, demonstrating that the auxiliary process admits a symmetry of
deformed spin algebra, explaining a peculiar bulk-cancellation mechanism, originally cast in terms of homogeneous cubic algebraic relations.
In spite the authors have succeeded in generalizing solutions via inclusion of ultra-local boundary coherent fields into the Hamiltonian, the central
algebraic condition, encapsulated in a suitable local operator-divergence form, has been employed without resorting to any underlying
(fundamental) algebraic principles. In particular no reference to the powerful tools of algebraic framework of the existing theory of quantum integrability
has been made or pointed out. At last, an undeniable evidence that constructed steady state solutions belong to the Yang-Baxter integrability
paradigm has been unfolded just recently \cite{PIP}, serving as prevailing motivation for the origin of this work.

The main goal of the present paper is mainly to elaborate on a direct relation of all presently known steady state solutions of anisotropic spin-$1/2$
Heisenberg spin chains to the formalism of quantum integrability, i.e. enlightening the connection to solutions of a famous Yang-Baxter equation associated to
quantum group structures. The presented approach allows for unified treatment of the models with integrable bulk interaction pertaining to
so-called \textit{fundamental integrable models}.

The paper is structured as follows. For the sake of completeness, we start in \Sref{sec:QG} by briefly presenting the construction of quantum groups via approach of Faddeev, Reshetikhin and Takhtajan, along with the introduction of related basic and essential concepts
from QISM. We proceed in \Sref{sec:lindblad} by outlining our setup for studying nonequilibrium states of quantum chains exhibiting ultra-local
boundary dissipation processes, which allow for elegant description of some non-trivial time-asymptotic states. In \Sref{sec:solutions} we present our main result, a general construction 
of nonequilibrium steady states of Markovian quantum evolution with unitary integrable bulk evolution and ultra-local boundary dissipation,
and apply it to simple models. In particular, a complete derivation of the steady state density operator for anisotropic spin-$1/2$ chain with unconstrained dissipator is presented in
a compact algebraic formulation. We demonstrate how the only compatible dissipator is the known maximally polarizing incoherent driving. Additionally, we supplement
 those results with analogous solutions for the isotropic higher-spin chains or, in other words, multicomponent particle-preserving quantum gases admitting
a global $SU(N)$ symmetry, using a restricted (primitive) set of Lindblad operators. In \Sref{sec:observables} we discuss a possibility of analytic evaluation of local physical observables
in terms of vertex operators imposed on the auxiliary spaces, establishing a relation between particle current density
and ratio of nonequilibrium partition functions pertaining to systems whose sizes differ for one particle.
We conclude by stressing out some intriguing connections to newly discovered quasi-local conserved operators and discuss some possible
futher improvements.

\section{Faddeev-Takhtajan-Reshetikhin construction of quantum groups}
\label{sec:QG}
One of indisputably most celebrated and influential results of the Leningrad's school of mathematical physics is
Faddeev's, Takhtajan's and Reshetikhin's (FRT) work on 'quantization' of Lie algebras~\cite{FRT}, which is formally speaking a realization of
algebraic objects arising as suitable continuous deformations of Lie-algebraic structures. These quantized algebras, simply
called \textit{quantum groups}, materialize as solutions of the celebrated \textit{Yang-Baxter equation}, and represents a cornerstone
of the quantum inverse scattering method.
Subsequently, those algebraic structures have been given more formal mathematical meaning independently by
Drinfeld~\cite{Drinfeld} and Jimbo~\cite{Jimbo85}.
In particular, quantum groups can be understood as non-trivial (so-called quasi-triangular) \textit{Hopf algebras}.

There exists an extensive literature on the subject~\cite{Majid,Klimyk97,Chaichian96,Jimbo1}.
Here, mainly for the sake of completeness, we shall follow the FRT formulation~\cite{FRT}. For our purpose, a quantum group is understood
as a one-parametric deformation of a universal enveloping algebra $\mathcal{U}(\mathfrak{g})$ for a given Lie algebra $\mathfrak{g}$.
We shall restrict ourselves entirely on $q$-deformations of algebra $\mathcal{U}(\mathfrak{sl}(N,\CC))$, merely for its importance in one-dimensional quantum models.

With the idea to motivate a rather abstract forthcoming discussion, we begin by introducing main objects and concepts of QISM.

\subsection{Brief overview: Quantum Inverse Scattering Method}
A principal goal of QISM is to establish a sufficient condition a (quantum) system has to be compliant with in order to posses countably
\textit{infinite number} of \textit{integrals of motion} (conserved charges). Imposing
periodic boundary conditions for a system consisting of $n$ copies of local $N$-level quantum spaces $\mathfrak{h}\cong \CC^{N}$,
i.e. $\mathfrak{H}_{s}=\mathfrak{h}_{1}\otimes \cdots \mathfrak{h}_{n}=\mathfrak{h}^{\otimes n}$,
a set of \textit{mutually commuting} local operators -- Hamiltonians $\{H_{m}\}$ from
$\mathfrak{H}_{s}$ -- arise from analytic series expansion of a parameter dependent \textit{quantum transfer operator} $\tau(\lambda)$,
at the so-called \textit{shift point} (or regular point) $\lambda_{0}$, as a consequence of its involutive property
\begin{equation}
[\tau(\lambda),\tau(\mu)]=0,\qquad \lambda,\mu \in \CC.
\end{equation}
Explicitly, a desired set of local conserved charges, with $H_{2}$ being interpreted as a Hamiltonian, is recovered upon taking logarithmic
derivatives of $\tau(\lambda)$ with respect to the complex \textit{spectral parameter} $\lambda$ around $\lambda_{0}$:
\begin{equation}
H_{m}\sim \left(\frac{d}{d\lambda}\right)^{m}\ln \tau(\lambda)|_{\lambda_{0}}.
\end{equation}
The transfer operator $\tau(\lambda)$ is canonically defined by means of a ``larger'' operator -- the \textit{monodromy matrix} ${\bf T}_{a}(\lambda)$,
as an operator in $\mathfrak{H}_{a}$  (being a Hilbert space associated to auxiliary degrees
of freedom) with elements from $\mathfrak{H}_{s}$ -- as a \textit{partial trace} over auxiliary
space $\mathfrak{H}_{a}$, $\tau(\lambda)=\Tr_{a}\bb{T}_{a}(\lambda)$. Throughout the paper we shall write boldface symbols whenever objects operate non-trivially on both, the auxiliary and the spin space.

The key property of monodromy matrices is the existence of a \textit{similarity transformation} $R(\lambda,\mu)$ which intertwines a product of two
monodromy matrices $\bb{T}(\lambda)\bb{T}(\mu)$ for different values of the spectral parameters, i.e. exchanging spectral parameters via
RTT equation,
\begin{equation}
R_{12}(\lambda,\mu)\bb{T}_{1}(\lambda)\bb{T}_{2}(\mu)=\bb{T}_{2}(\mu)\bb{T}_{1}(\lambda)R_{12}(\lambda,\mu).
\label{FCR}
\end{equation}
%or in component notation (using Einstein convention)
%\begin{equation}
%R^{ab}_{cd}(\lambda,\mu)T^{c}_{e}(\lambda)T^{d}_{f}(\mu)=T^{b}_{d}(\mu)T^{a}_{c}(\lambda)R^{cd}_{ef}(\lambda,\mu).
%\label{RTT_algebra}
%\end{equation}
The latter can be thought of as a definition of an abstract homogeneous associative \textit{quadratic algebra} composed of non-commuting elements
$\{ T^{i}_{j}(\lambda)\}$, with $R$-matrix prescribing algebra's structure constants. In equation \eref{FCR} we emploed a standard index notation
where subscript indices refer to auxiliary spaces in which $R$ and $\bb{T}$ matrices operate non-trivially, namely $R$-matrix is a linear map in
$\mathfrak{H}_{1}\otimes \mathfrak{H}_{2}$ (thus its elements are \textit{scalars} with respect to $\mathfrak{H}_{s}$),
and $\bb{T}_{j}(\lambda)$ operates non-trivially in $\mathfrak{H}_{j}\otimes \mathfrak{H}_{s}$. The intertwining $R$-matrix must satisfy additional compatibility-type condition,
\begin{equation}
R_{12}(\lambda,\mu)R_{13}(\lambda,\eta)R_{23}(\mu,\eta)=R_{23}(\mu,\eta)R_{13}(\lambda,\eta)R_{12}(\lambda,\mu),
\label{YBE}
\end{equation}
known as the \textit{Yang-Baxter equation} (YBE), in order to ensure the associativity of the RTT algebra \eref{FCR}.
Notice how, in contradistinction to \eref{FCR}, the condition \eref{YBE} is imposed over three-fold product of auxiliary spaces
$\mathfrak{H}_{1}\otimes \mathfrak{H}_{2}\otimes \mathfrak{H}_{3}$, with $R$-matrices (with elements in $\CC$) equipped with
subscript indices pertaining to spaces where they operate non-trivially, i.e. \textit{non-identically}.

The YBE can be regarded as a generalization of the permutation group. Introducing a permutation map $\Pi$ of two tensor factors,
\begin{equation}
\Pi(a\otimes b)=P(a\otimes b)P=b\otimes a,
\end{equation}
with $a\otimes b$ from $\mathfrak{H}_{1}\otimes \mathfrak{H}_{2}$,
we essentially obtain a \textit{constant} solution $P_{12}P_{13}P_{23}=P_{23}P_{13}P_{12}$.
Equivalently, one can define a \textit{braid} $R$-matrix, $\PR(\lambda,\mu):=PR(\lambda,\mu)$,
which fulfills a braid-associativity condition,
\begin{equation}
\PR_{12}(\lambda,\mu)\PR_{23}(\mu,\eta)\PR_{12}(\lambda,\mu)=\PR_{23}(\mu,\eta)\PR_{12}(\lambda,\mu)\PR_{23}(\mu,\eta).
\end{equation}
The RTT equation \eref{FCR} is in fact, due to locality principle, i.e. factorization of space $\mathfrak{H}_{s}$ on local physical spaces $\mathfrak{h}_{j}$,
implied by existence of a \textit{local unit}, i.e. an operator $\bb{L}_{a,j}(\lambda)\equiv \bb{L}_{j}(\lambda)$ over
$\mathfrak{H}_{a}\otimes\mathfrak{h}_{j}$, usually referred to as \textit{quantum Lax operator},
by means of a \textit{spatially-ordered product} of the form
\begin{equation}
\bb{T}_{a}(\lambda)=\bb{L}_{1}(\lambda)\bb{L}_{2}(\lambda)\cdots \bb{L}_{n}(\lambda)=:\prod_{j=1}^{n}\bb{L}_{j}(\lambda).
\end{equation}
Here all operators $\bb{L}_{j}(\lambda)$ are over $n$-fold product physical space $\mathfrak{H}_{s}$,
with matrix elements from $\mathfrak{H}_{a}$, and index $j$ denotes local physical spaces $\mathfrak{h}_{j}$
where $\bb{L}_{j}$ operate non-trivially.

%We however prefer an alternative notation in this paper, in which the monodromy matrix is expressed as
%\begin{equation}
%T(\lambda)=L(\lambda)\otimes_{s} L(\lambda)\otimes_{s}\cdots \otimes_{s}L(\lambda)=:L(\lambda)^{\otimes_{s} n},
%\end{equation}
%where $\otimes_{s}$ designates a partial tensor product operation over $\mathfrak{H}_{a}\otimes \mathfrak{h}\otimes \mathfrak{h}$,
%namely a tensor product with respect to two copies of local spin spaces $\mathfrak{h}$ and ordinary matrix multiplication with respect to
%one copy of auxiliary space $\mathfrak{H}_{a}$. The counterpart operator $\otimes_{a}$ is then naturally employed in the local version of
%the fundamental commutation relation[ref] when expressing similarity of two parameter-dependent Lax matrices,
%\begin{equation}
%L(\lambda)\otimes_{a}L(\mu)=R(\lambda,\mu)^{-1}(L(\mu)\otimes_{a}L(\lambda))R(\lambda,\mu)
%\end{equation}

With help of the property \eref{FCR}, it is easy to establish the commutativity of transfer matrices by tracing over spaces
$\mathfrak{H}_{1},\mathfrak{H}_{2}$, by virtue of
\begin{eqnarray}
\tau(\lambda)\tau(\mu)&=\Tr_{12}(\bb{T}_{1}(\lambda)\bb{T}_{2}(\mu))\nonumber \\
&=\Tr_{12}(\PR_{12}(\lambda,\mu)(\bb{T}_{1}(\lambda)\bb{T}_{2}(\mu))\PR_{12}(\lambda,\mu)^{-1})\nonumber \\
&=\Tr_{12}(\bb{T}_{1}(\mu)\bb{T}_{2}(\lambda))=\tau(\mu)\tau(\lambda),
\end{eqnarray}
by merely accounting for the definition of the monodromy matrix $\bb{T}_{a}(\lambda)$ and associativity of matrix multiplication.

\subsection{The universal $R$-matrix and $L$-matrices}
Pursuing the FRT approach~\cite{FRT} we now justify the role of the RTT equation in the language of quantum groups,
i.e. a quantizations (deformations) of Lie algebraic structures.
Imagine an associative quadratic algebra $\mathcal{A}(R)$ (over $\CC$) generated by a unit $1$ and a set of $N^{2}$ elements
$\{ T^{i}_{j}\}$, $i,j=1,\ldots N$, with $R$-matrix $R$, obeying \eref{FCR}. The algebra admits a \textit{bialgebra} structure,
with \textit{coproduct} $\Delta:\mathcal{A}\rightarrow \mathcal{A}\otimes \mathcal{A}$ prescribed as
\begin{equation}
\Delta(1)=1\otimes 1,\quad \Delta(T^{i}_{j})=\sum_{k}T^{i}_{k}\otimes T^{k}_{j},
\end{equation}
and \textit{counit} $\epsilon:\mathcal{A}\rightarrow \CC$, $\epsilon(T^{i}_{j})=\delta^{i}_{j}$.
Denoting $\mathcal{A}^{*}$ as a \textit{dual space} of $\mathcal{A}$, $\mathcal{A}^{*}$ is generated by $1^{\prime}$ and $\{l^{\pm}_{ij}\}$,
such that $1^{\prime}(T^{i}_{j})=\delta_{i,j}$.  A coproduct in $\mathcal{A}$ is induced by the product (multiplication) in $\mathcal{A}^{*}$,
namely for two elements $l_{1},l_{2}\in \mathcal{A}^{*}$ and $a\in \mathcal{A}$ we have in particular
\begin{equation}
(l_{1}l_{2},a)\equiv l_{1}l_{2}(a)=(l_{1}\otimes l_{2})(\Delta(a)).
\end{equation}
By focusing on $\mathfrak{g}\cong \mathfrak{gl}(N,\CC)$ and its $q$-deformation,
one can introduce two Borel matrices-functionals $\bb{L}^{\pm}$ with elements from %\in M_{n}(\mathcal{A}^{*})
$\mathcal{A}^{*}$, i.e. $[\bb{L}^{\pm}]^{i}_{j}=l^{\pm}_{ij}$,
and express this \textit{duality} compactly as
\begin{equation}
(\bb{L}^{\pm},\bb{T})=R^{\pm}_{q},\qquad (1^{\prime},\bb{T})=\mathds{1},
\end{equation}
via Borel (invertible) matrices $R^{\pm}_{q}$,
\begin{equation}
R^{+}_{q}=PR_{q}P,\qquad R_{q}^{-}=R_{q}^{-1},
\end{equation}
prescribed by means of the lower-triagular matrix $R_{q}$ operating in $\CC^{N}\otimes \CC^{N}$,
\begin{equation}
R_{q}=\sum_{i\neq j=1}^{N}E^{ii}\otimes E^{jj}+q\sum_{i=1}^{N}E^{ii}\otimes E^{ii}+(q-q^{-1})\sum_{1\leq j\le i\leq N}E^{ij}\otimes E^{ij}.
\label{R_glN}
\end{equation}
The independent set of equations, provided by (omitting $q$ label for the moment, and using index notation)
\begin{eqnarray}
R^{+}\bb{L}^{\pm}_{1}\bb{L}^{\pm}_{2}=\bb{L}^{\pm}_{2}\bb{L}^{\pm}_{1}R^{+},\nonumber \\
R^{\pm}\bb{L}^{\pm}_{1}\bb{L}^{\mp}_{2}=\bb{L}^{\mp}_{2}\bb{L}^{\pm}_{1}R^{\pm},
\label{FRT_equations}
\end{eqnarray}
serves as an abstract definition of a $q$-deformation of a 'classical' Lie algebra $\mathfrak{gl}(N,\CC)$,
which shall be facilitated in our construction of a steady state density operator in \Sref{sec:solutions}.
The induced coproduct in the dual bialgebra reads
\begin{equation}
\delta(1')=1^{\prime}\otimes 1^{\prime},\quad \delta(l^{\pm}_{ij})=\sum_{k}l^{\pm}_{ik}\otimes l^{\pm}_{kj}.
\end{equation}

\paragraph{Universal $\mathcal{R}$-matrix.}
Recalling \eref{FCR}, the quadratic algebra of monodromy matrix elements $\{T^{i}_{j}\}$, written out component-wise using Einstein convention,
\begin{equation}
R^{ij}_{ab}T^{a}_{k}T^{b}_{l}=T^{j}_{b}T^{i}_{a}R^{ab}_{kl},
\label{RTT_elements}
\end{equation}
is prescribed via $R$-matrix, a constant solution of aforementioned YBE \eref{YBE}. We may formally
express the \textit{non-cocommutativity} of the coproduct in bialgebra $\mathcal{A}$ via identification,
\begin{equation}
R^{ij}_{kl}\equiv \mathcal{R}(T^{i}_{j}\otimes T^{k}_{l}),
\end{equation}
by means of a universal $\mathcal{R}$-matrix, governing the similarity of the coproduct and an opposite coproduct,
\begin{equation}
\mathcal{R}\Delta(a)=(\Pi\circ \Delta)(a)\mathcal{R}.
\label{quasitriangular}
\end{equation}
The non-trivial condition of the latter type endows a Hopf algebra with a \textit{quasi-triangular} structure.
The associativity of the coproduct over triple-product algebra $\mathcal{A}\otimes\mathcal{A}\otimes \mathcal{A}$ requires
the two expressions,
\begin{equation}
(\Delta\otimes \mathds{1})\mathcal{R}=\mathcal{R}_{13}\mathcal{R}_{23},\qquad (\mathds{1}\otimes \Delta)\mathcal{R}=\mathcal{R}_{13}\mathcal{R}_{12}
\end{equation}
to be equivalent, hence
\begin{equation}
\mathcal{R}_{12}\mathcal{R}_{13}\mathcal{R}_{23}=\mathcal{R}_{23}\mathcal{R}_{13}\mathcal{R}_{12}.
\label{universal_YBE}
\end{equation}
Thus, we have finally established the connection of a quantum group to the (parameter independent) Yang-Baxter equation.
Evaluating the universal element $\mathcal{R}_{12}$ from \eref{universal_YBE} for our choice of deformed Lie algebra $\mathfrak{gl}(N,\CC)$ in $\CC^{N}\otimes \CC^{N}$
explicitly yields \eref{R_glN}. Notice that in QISM we employ parameter-dependent solutions \eref{YBE} which are formally obtained via
\textit{Baxterization}~\cite{Jones} of quantum groups objects.

The reader interested in more detailed and concise presentation of quantum groups is referred to e.g. \cite{Majid,FRT}.

\paragraph{Example: deformation of $\mathfrak{sl}(2,\CC)$ algebra.}
The simplest example is to consider an algebra of spin generators, i.e. the $\mathfrak{sl}(2,\CC)$-triple
$\{S^{\pm},S^{z}\}$ with canonical (Lie algebraic) commutation relation,
\begin{equation}
[S^{z},S^{\pm}]=\pm S^{\pm},\quad [S^{+},S^{-}]=2S^{z}.
\end{equation}
The associated universal enveloping algebra $\mathcal{U}(\mathfrak{sl}(2,\CC))$ is a tensor algebra freely
generated by $\{S^{z},S^{\pm}\}$, i.e. a tensor algebra of words formed from the algebra generators, modulo their commutation relations.
One can now introduce the $q$-deformed universal enveloping algebra $\Uqsl{2,\CC}$, with deformation parameter $q\in \CC$,
adopting standard $q$-calculus definition for the (deformed) $q$-number,
\begin{equation}
[x]_{q}:=\frac{q^{x}-q^{-x}}{q-q^{-1}},
\label{qnumber}
\end{equation}
as a quantum group with elements generated by identity and four quantized generators
$\{S^{\pm},K^{\pm}\}$, where $K^{\pm}=q^{\pm S^{z}}$,
subjected to the defining commutation relations
\begin{equation}
[S^{+},S^{-}]=[2S^{z}]_{q}=\frac{(K^{+})^{2}-(K^{-})^{2}}{q-q^{-1}},\quad
K^{\pm}S^{\pm}=q^{\pm 1}S^{\pm}K^{\pm},
\label{deformed_relations}
\end{equation}
with operator $q$-analogues defined via analytic series expansion in analogy to \eref{qnumber}.
It is noteworthy that the undeformed limit corresponds to the value of deformation parameter $q=1$, however, to properly restore 'classical' Lie algebra one
has to take the differential form of the \eref{deformed_relations}.
We remark that this particular form of 'quantization' is of main relevance when considering \textit{axial deformations} of $SU(N)$-symmetric
quantum chain models. Subsequently in our discussion, we shall restrict ourselves to $q$ parameter from the \textit{unit circle},
\begin{equation}
q=e^{\ii \gamma},\quad \gamma\in \mathbb{R}.
\end{equation}

In a language of FRT algebra, a quasi-triangular structure for the $\mathcal{U}_{q}(\mathfrak{sl}(N,\CC))$ is explicitly prescribed by the
universal element
\begin{equation}
\mathcal{R}=q^{2(S^{z}\otimes S^{z})}
\sum_{k=0}^{\infty}\frac{(1-q^{-2})^{k}}{[k]!}\left(q^{S^{z}}S^{+}\otimes q^{-S^{z}}S^{-}\right)^{k}q^{k(k-1)/2}.
\label{universal_R}
\end{equation}
Evaluating this expression in the fundamental (spin-$1/2$) representation, and accounting that generators $S^{\pm}$ square to zero,
immediately yields (up to a prefactor) the previously stated result \eref{R_glN}. Moreover, the explicit form for
$L$-matrices \eref{FRT_L_explicit} is expressed in a canonical way element-wise as
\begin{equation}
(L^{+}_{kl})^{i}_{j}=R^{ik}_{jl},\quad (L^{-}_{kl})^{i}_{j}=(R^{-1})^{ki}_{lj}.
\end{equation}
In words, $L$-matrices are obtained from the $R$-matrix by associating block elements of the second tensor factor with algebra generators.
%$\rho_{1}(L^{+}_{2})=R$ and $\rho_{1}(L^{-}_{2})=R^{-1}_{21}$.
This in turn leads to the following pair of $L$-operators (with elements proportional to the generators of $\Uqsl{2,\CC}$),
\begin{equation}
\bb{L}^{+}=\pmatrix{
K & (q-q^{-1})S^{-} \cr
0 & K^{-1}
},\qquad
\bb{L}^{-}=\pmatrix{
K^{-1} & 0 \cr
-(q-q^{-1})S^{+} & K
},
\label{FRT_L_explicit}
\end{equation}
and Borel components of the $R$-matrix,
\begin{equation}
R^{+}=\pmatrix{
q & & & \cr
& 1 & q-q^{-1} & \cr
& & 1 & \cr
& & & q
},\qquad R^{-}=P(R^{+})^{-1}P.
\label{FRT_R_explicit}
\end{equation}
One can now readily check that \eref{FRT_equations} are consistent with the defining relations of $\Uqsl{2,\CC}$ \eref{deformed_relations}.
We therefore have a convenient algebraic expression for quantum group relations in an abstract fashion, namely the functions of generators
appearing in matrix elements can readily be taken from any representation,
i.e. the space in which the spin generators operate can be chosen arbitrarily.

As we demonstrate later on \ref{sec:solutions}, when construction of solutions to our nonequilibrium problem is presented,
this freedom will be of crucial importance.
In other words, the FRT algebra \eref{FRT_equations} materializes as a prerequisite bulk cancellation mechanism - a sort of quantum version
of an analogous condition in the context of classical exclusion processes~\cite{Derrida}.

Ultimately, the complete quantum group structure of the $\Uqsl{2,\CC}$ is explicitly provided by Hopf algebra co-structures,
namely the coproduct $\Delta$ and counit $\epsilon$,
\begin{eqnarray}
\Delta(S^{\pm})&=S^{\pm}\otimes K^{+}+K^{-}\otimes S^{\pm},\quad \Delta(K^{\pm})=K^{\pm}\otimes K^{\pm},\\
\epsilon(S^{\pm})&=0,\quad \epsilon(K^{\pm})=1,
\end{eqnarray}
along with the antipode map $\zeta$, prescribing an inverse operation by virtue of $\zeta(\bb{T})=\bb{T}^{-1}$,
\begin{equation}
\zeta(S^{\pm})=-q^{\pm 1}S^{\pm},\quad \zeta(K^{\pm})=K^{\mp},
\end{equation}
which completes a bialgebra to a Hopf algebra.
At last, a quasi-triangular structure is given by the $R$-matrix in a sense of \eref{quasitriangular}.

\section{Lindblad master equation with boundary dissipation}
\label{sec:lindblad}
% contraction with respect to lowest-weight state (explanation why - lowest numbers of terms to match)
Let us now switch gears and discuss a seemingly unrelated problem of dissipative evolution of a many-body \textit{open} quantum system.
We shall introduce a setup of \textit{boundary-driven} Markovian open quantum system and present recent findings of steady states for the
Heisenberg $XXZ$ spin-$1/2$ chain, to prepare a terrain for the general construction of steady states by means of
integrability structures presented in \Sref{sec:QG}.
\subsection{Steady state solution of Markovian quantum master equation}
Our aim is to address an idea of constructing analytic exact steady states of nonequilibrium quantum many-body evolution for some
paradigmatic one-dimensional models of strongly interacting particles. By facilitating a master equation description, i.e. writing
the most general \textit{completely positive trace-preserving map}~\cite{Lin76}
for the system's density matrix with time-independent generator, and focusing solely on the steady state density operators,
we seek for fixed-point solutions of the Markovian flow,
\begin{equation}
\rho(t)=\VV(t)\rho(0),\qquad \VV(t)=e^{\LL t},
\end{equation}
with the \textit{infinitesimal generator} $\LL$ of the \textit{Lindblad form},
\begin{equation}
\frac{d}{dt}\rho(t)=\LL \rho(t),\quad \LL(\rho)=-\ii \adH(\rho)+\DD(\rho),
\label{Lindblad_equation}
\end{equation}
with linear super-operator maps
\begin{equation}
\adH(\rho):=[H,\rho],\qquad
\DD(\rho):=\sum_{\mu}A_{\mu}\rho A^{\dagger}_{\mu}-\frac{1}{2}\{A^{\dagger}_{\mu}A_{\mu},\rho\},
\end{equation}
where $\{A,B\}:=AB+BA$ denotes the anti-commutator.
Operators from the set $\{A_{\mu}\}$ are called the \textit{Lindblad operators}, modeling \textit{incoherent} (non-unitary) part of quantum evolution.
In our setup, their role will mainly be to establish a gradient of a 'chemical potential', i.e. to introduce forces inducing current-carrying states and thus
allowing to study genuine far-from-equilibrium situations. At this point we would like to emphasize that using Lindblad operators to model a sort of realistic
reservoirs or thermal baths is out of scope in our setting, as essentially we only care about providing simple enough conditions for analytic treatment to force a system out of equilibrium.
Consequently, one has to understand that the steady state under discussion will be generically far from the linear-response regime.
To this end, we adopt a setup where dissipative processes affect only the first and the last particle in a chain, namely we take the
dissipator to be of the form
\begin{equation}
\DD(\rho)=\DD_{\rm L}(\rho)+\DD_{\rm R}(\rho),
\end{equation}
with dissipation super-operator $\DD_{\rm L,R}$ operating non-trivially only in the first/last local quantum space,
$\mathfrak{h}_{1}$ and $\mathfrak{h}_{n}$, respectively.
The nonequilibrium steady state density operator (NESS) is therefore a \textit{fixed-point solution}
$\rho_{\infty}$ of the \eref{Lindblad_equation}, i.e. defined by $\LL(\rho_{\infty})=0$, or
\begin{equation}
\ii \adH(\rho_{\infty})=\DD_{\rm L}(\rho_{\infty})+\DD_{\rm R}(\rho_{\infty}).
\label{Lindblad_fixed}
\end{equation}

\subsection{Exact solution for a boundary-driven anisotropic Heisenberg spin-$1/2$ chain}
In this subsection we briefly discuss the steady state for the anisotropic Heisenberg spin-$1/2$ chain found in \cite{PRL106,PRL107}.
The solution to \eref{Lindblad_fixed} with a global Hamiltonian $H$ operating over $n$-body Hilbert space
$\mathfrak{H}_{s}\cong (\CC^{2})^{\otimes n}$
\begin{equation}
H^{XXZ}=\sum_{j=1}^{n-1}h^{XXZ}_{j,j+1},\quad h_{j,j+1}=2\sigma^{+}_{j}\sigma^{-}_{j+1}+2\sigma^{-}_{j}\sigma^{+}_{j+1}+
\cos{(\gamma)}\sigma^{z}_{j}\sigma^{z}_{j+1},
\label{Hamiltonian_XXZ}
\end{equation}
where $\cos{(\gamma)}\in[-1,1]$ defines the \textit{anisotropy parameter}, and \textit{maximally polarizing} left/right channels, given by
\begin{equation}
A_{1}=\sqrt{\Gamma}\sigma^{+}_{1},\qquad A_{n}=\sqrt{\Gamma}\sigma^{-}_{n},
\label{Tomaz_dissipators}
\end{equation}
with coupling strength parameter $\Gamma \in \mathbb{R}$,
can be cast as \textit{Cholesky-type factorized} (non-normalized) density matrix $\rho_{\infty}=S_{n}S_{n}^{\dagger}$.
The 'Cholesky' factor $S_n$ is defined in a matrix-product operator (MPO) form expanded in the
standard (Weyl) basis in $\mathfrak{H}_{s}$ using binary vectors $\underline{\alpha}=(\alpha_{1},\ldots,\alpha_{n})$ and
$\underline{\beta}=(\beta_{1},\ldots,\beta_{n})$ for $\alpha_{j},\beta_{j}\in \{0,1\}$,
\begin{equation}
S_{n}=\sum_{\underline{\alpha}}\sum_{\underline{\beta}}
\bra{l}L^{\alpha_{1}\beta_{1}}L^{\alpha_{2}\beta_{2}}\cdots L^{\alpha_{n}\beta_{n}}\ket{r}
\bigotimes_{j=1}^{n} E^{\alpha_{j}\beta_{j}}_{j},
\label{MPO_S}
\end{equation}
with a set of auxiliary matrices $\{L^{ij}\}$.
Throughout the paper we interchangeably use two complete basis sets of linear operators on a local qubit space $\mathfrak{h}\cong \CC^{2}$,
the canonical Pauli matrices
\begin{equation}
\sigma^{+}=\pmatrix{0 & 1 \cr 0 & 0},\quad \sigma^{-}=\pmatrix{0 & 0 \cr 1 & 0},\quad \sigma^{z}=\pmatrix{1 & 0 \cr 0 & -1},
\end{equation}
along with $\sigma^{0}\equiv \mathds{1}_{2}$, and the standard unit matrices (Weyl basis) $\{E^{ij}\}$, for $i,j=0,1$, satisfying
$E^{ij}E^{kl}=E^{il}\delta_{jk}$ and $\mathfrak{gl}(n,\CC)$ (ultra-local) commutation relations
\begin{equation}
\label{fundamental_comm_relations}
[E^{ij}_{m},E^{kl}_{p}]=(E^{il}_{m}\delta_{jk}-E_{m}^{kj}\delta_{il})\delta_{mp}.
\end{equation}
The strategy to find a steady state density matrix is based on a compact presentation via homogeneous MPO of the form \eref{MPO_S}, by means of spatially-ordered product of
local matrices ${\bf L}_j(\Gamma)$,
\begin{equation}
S_{n}(\Gamma)=\bra{l}\bb{L}_{1}(\Gamma)\bb{L}_{2}(\Gamma)\cdots \bb{L}_{n}(\Gamma)\ket{r}=\bra{l}\prod_{j=1}^{n}\bb{L}_{j}(\Gamma)\ket{r}.
\label{ansatz_S}
\end{equation}
The elements of ${\bf L}_j$ are again non-commuting matrices acting on an auxiliary space
\begin{equation}
\bb{L}_{j}(\Gamma)=\sum_{k,l=1}^{2}E^{kl}_{j}\otimes L^{lk}(\Gamma)=\pmatrix{
L^{11}(\Gamma) & L^{21}(\Gamma) \cr
L^{12}(\Gamma) & L^{22}(\Gamma) \cr
}_{j}.
\end{equation}
An ansatz of this type has been proposed in~\cite{KPS} and is in fact inspired directly from analogy to known matrix-product realizations
of partition functions for classical asymmetric exclusion process~\cite{Derrida,Schutz,Aneva},
lifted to quantum spaces (see also ~\cite{Temme}).

%We shall call $L$-matrix the \textit{Lax matrix}. In this regard, our factor operator $S_{n}$ from $\mathfrak{H}_{s}$ can be interpreted as the $T^{l}_{r}$
%matrix element of the corresponding monodromy matrix $\bb{T}(\Gamma)=\bb{L}_{1}(\Gamma)\cdots \bb{L}_{n}(\Gamma)$,
%although at this stage this may seem a slight abuse of terminology, since there is no obvious reason
%that any contraction (not necessarily a partial trace over $\mathfrak{H}_{a}$) would yield a quantum transfer matrix.
%We shall however return to this point later on.

Notice that the solution of \eref{Lindblad_fixed}, with dissipators acting ultra-locally only on the boundary sites, imposes a condition
on the action of the adjoint global Hamiltonian,
\begin{equation}
\ii \adH(S_{n})=\sum_{k}\left(w^{k}_{L}\sigma^{k}_{1}W^{k}_{R}-w^{k}_{R}W^{k}_{L}\sigma^{k}_{n}\right),
\label{adH_physical}
\end{equation}
with $W_{\rm L}^{k},W_{\rm R}^{k}$ operating in $\mathfrak{H}_{[2,n]}$ and $\mathfrak{H}_{[1,n-1]}$,
respectively. Operators from \eref{adH_physical} are essentially MPOs with the same local unit ${\bf L}_m$, only contracted differently.
In other words, $\adH$ has to non-trivially modify the density operator only in the boundary physical spaces. Consequently, a \textit{sufficient} requirement
to ensure this property at the level of MPO is to satisfy the following form of the local operator-divergence condition with respect to
interaction density $h_{j,j+1}$,
\begin{equation}
[h_{j,j+1},\bb{L}_{j}\bb{L}_{j+1}]=\bb{B}_{j}\bb{L}_{j+1} - \bb{L}_{j}\bb{B}_{j+1}.
\label{Sutherland_KPS}
\end{equation}
After multiplying by $\prod_{m=1}^{j-1}\bb{L}_{m}$ from the left and by $\prod_{m=j+2}^{n}\bb{L}_{m}$ from the right,
and contracting with respect to given left and right auxiliary boundary vectors, one arrives at the equation \eref{adH_physical}.
For example, the choice \eref{Tomaz_dissipators}, in compliance with reference~\cite{PRL107},
% compliant with the vacuum boundary states $\bra{l}\rightarrow \bra{0}$ and $\ket{r}\rightarrow\ket{0}$, and
yields \textit{non-zero} coefficients $w^{L}_{z}=-w^{R}_{z}=2\Gamma$, corresponding to $W^{L}_{z}=W^{R}_{z}=S_{n-1}$.

An explicit form of local matrix $\bb{L}$ and boundary auxiliary matrix $\bb{B}$ can be found in~\cite{KPS},
implemented by a set of $\Uqsl{2,\CC}$ generators acting in \textit{infinitely dimensional} auxiliary Hilbert space.
Nonetheless, no insight has been presented on why matrices with deformed symmetry are of fundamental interest in such setup and
how solutions of \eref{Sutherland_KPS} emerge from more fundamental (algebraic) principles. The purpose of forthcoming discussion
is therefore to make use of the presented formalism of quantum groups to (i) generate solutions of \eref{Sutherland_KPS} pertaining
to fundamental integrable models and derive a canonical representation of solutions with deformed symmetries, and (ii) to formulate and solve the
a system of compatibility (boundary) equations for a class of $SU(N)$ invariant models.

\section{Exact matrix-product form solutions for fundamental spin chains}
\label{sec:solutions}

Despite original there has been no transparent interrelation between presented NESS solutions and the existing theory of quantum integrability,
the algebraic formulation via objects with deformed continuous symmetry in \cite{KPS} has given a first hint in this direction.
Shortly after, a remarkable transfer-operator property has been revealed~\cite{PIP} upon noticing that the operator
$S_{n}(\Gamma)=T^{0}_{0}(\Gamma)$, defining the solution of the maximally-driven anisotropic Heisenberg chain, with the dissipation coupling
strength parameter $\Gamma$, forms a \textit{commuting family},
\begin{equation}
[S_{n}(\Gamma_1),S_{n}(\Gamma_2)]=0,\qquad \Gamma_{1,2} \in \CC.
\end{equation}
This finding, which called for the existence of the underlying Yang-Baxter structure, has been explained recently  by means of an
\textit{infinite-dimensional} $R$-matrix \cite{PIP}. 

In this section we unveil the connection between the QISM and the steady state solutions for boundary-driven dissipative one-dimensional models within
the framework introduced in \Sref{sec:lindblad}, and present a simple rigorous construction of NESS by using elementary ingredients
of the FRT approach from \Sref{sec:QG}. We furthermore elaborate on the proposed Cholesky-type factorization of the density matrix
and the role of infinite dimensional representations (Verma modules) for the Lax matrix and their importance for the construction of
the steady states. Finally, the procedure is presented on simple examples.

%In this section we show that underlying integrability of boundary-driven dissipative one-dimensional models introduced in \Sref{sec:lindblad} enables a simple construction of 
%NESS by using elementary concepts of the QISM discussed in \Sref{sec:QG}. We then elaborate on the proposed Cholesky-type factorization of the density matrix and the role of infinite dimensional
%representations (Verma modules) of the Lax matrix in the construction of the steady states. Finally, the procedure is presented on simple examples.

% mention here that later it became clear that these object have been previously constructed in different contexts, i.e. CFT and QCD.

%Another indication can be attributed to the fact that we are seeking for (at least) one-parametric family of NESS density operators,
%determined by continuous parameters defining the dissipator. Therefore, having employed the Lax operator representation, solution of
%operator divergence equation \eref{Sutherland_KPS} has to allow for sufficient freedom, namely solutions could not simply be determined
%in terms of constant matrices by fixing the dimension of auxiliary space $\mathfrak{H}_{a}$, as we would lose a continuous
%freedom we need in order to satisfy the conditions imposed by the boundary equations \eref{boundary_compact}.

\subsection{Sutherland equation: local divergence condition}
First, let us demonstrate how solutions of \eref{Sutherland_KPS} emerge from an \textit{infinitesimal form} of fundamental solutions of the YBE \eref{YBE}.
Essentially, the equation \eref{Sutherland_KPS} is nothing but Sutherland's local condition which is sufficient for establishing commutativity
between the Hamiltonian with local density $h$ and a transfer matrix obtained as a partial trace over monodromy matrix for a given Lax matrix $\bb{L}$,
assuming periodic boundary conditions~\cite{Sutherland}. The same fact has been elaborated on in Sklyanin's lecture notes~\cite{Sklyanin}.

Our starting point is a parametrized local Yang-Baxter equation -- the so-called $RLL$ relation --  of the \textit{difference form},
defined in two-fold product of auxiliary spaces $\mathfrak{H}_{1}\otimes \mathfrak{H}_{2}$,
\begin{equation}
R_{12}(\lambda-\mu)\bb{L}_{1}(\lambda)\bb{L}_{2}(\mu)=\bb{L}_{2}(\mu)\bb{L}_{1}(\lambda)R_{12}(\lambda-\mu),
\label{local_YBE}
\end{equation}
obtained by Baxterization~\cite{Jones} of the objects from \eref{FRT_equations} with spectral parameter $\lambda$,
via variable $x=q^{-\ii \lambda}=e^{\gamma \lambda}$,
\begin{eqnarray}
R^q_{12}(\lambda)&=xR_{12}^{+}-x^{-1}R_{12}^{-},\\
\bb{L}_{j}^q(\lambda)&=x\bb{L}_{j}^{+}-x^{-1}\bb{L}_{j}^{-},\quad j=1,2.
\end{eqnarray}
Note that choosing $\lambda=\mu$ (or equivalently $x=1$) in the equation \eref{local_YBE} defines a \textit{shift point},
where the $R$-matrix becomes proportional to the
permutation operator in $\mathfrak{H}_{1}\otimes \mathfrak{H}_{2}$,
\begin{equation}
\label{shift_R}
R_{12}^{q}(0)=(q-q^{-1})P_{12},
\label{regularity}
\end{equation}
implying $\PR^q(0)=(q-q^{-1})\mathds{1}$.
Taking the derivative of \eref{local_YBE} with respect to $\lambda$ at $\lambda=\mu$, and applying permutation operator from the left
results in the \textit{infinitesimal} $RLL$-relation,
\begin{eqnarray}
\small
[\partial_{\lambda}\PR^{q}_{12}(\lambda)|_ {\lambda=0},\bb{L}_{1}(\lambda)\bb{L}_{2}(\lambda)]=
&-\PR^{q}_{12}(0)(\partial_{\lambda}\bb{L}_{1}(\lambda))\bb{L}_{2}(\lambda)\nonumber \\
&+\bb{L}_{1}(\lambda)(\partial_{\lambda}\bb{L}_{2}(\lambda))\PR^{q}_{12}(0),
\end{eqnarray}
which is after using \eref{shift_R} equivalent to the \textit{Sutherland equation}
(up to trivial rescaling of operators $\PR^{q}_{12}(\lambda)$ and $\bb{L}_{j}(\lambda)$)
\begin{equation}
[h_{12},\bb{L}_{1}(\lambda)\bb{L}_{2}(\lambda)]=\bb{B}_{1}(\lambda)\bb{L}_{2}(\lambda)-\bb{L}_{1}(\lambda)\bb{B}_{2}(\lambda).
\label{Sutherland_equation}
\end{equation}

Finally, a comparison with the previously stated form of the Sutherland equation \eref{Sutherland_KPS} indicates that we need to formally
identify auxiliary spaces $\mathfrak{H}_{1}$ and $\mathfrak{H}_{2}$ with local physical spaces, $\mathfrak{h}_{j}$ and
$\mathfrak{h}_{j+1}$, respectively, whereas matrix elements of $L$-operators, i.e. the spin-algebra generators now operate in the 'third space'
which we interpret as the auxiliary space $\mathfrak{H}_{a}$.  Thus, we can identify local auxiliary matrices ${\bf L}$ in the
ansatz \eref{MPO_S} with Lax operators of the corresponding integrable bulk Hamiltonian, automatically satisfying the local divergence condition \eref{Sutherland_KPS}.

In the next step we shall construct an ansatz for the NESS by using the bulk cancellation properties for the MPO $S_{n}$.

\subsection{Cholesky-factorized form and boundary equations}
As we have seen, an ansatz of the NESS density matrix can be formulated as Cholesky-type\footnote{In order to deal with standard
Cholesky decomposition, $S_{n}$ has to be in addition a triangular operator in quantum many-body
basis. However, as a consequence of non-unitarity of representations for auxiliary spin and vacuum (lowest weight) contraction, this is
always the case for solutions we discuss in this paper.} decomposition with an MPO factor $S_{n}$.
Despite at least intuitively such form appears to be a good candidate as it represents a manifestly positive operator,
it seems difficult to justify it in a rigorous way\footnote{It is perhaps noteworthy that solutions of a similar type,
however not strictly of Cholesky-form, which lack transfer matrix property,
have already been constructed for anisotropic Heisenberg $s=1/2$ model with \textit{asymmetric} boundary incoherent driving via model-specific
generalized ansatz~\cite{asymmetric}.}.
One could for instance propose a simpler ansatz, where NESS operator $\rho_{\infty}$ is sought as a \textit{linear}
expression in MPOs, i.e. as $\rho_{\infty}\sim S_{n}\pm S_{n}^{\dagger}$ with $S$-operator \eref{ansatz_S}.
Nevertheless, for spin-$1/2$ chains, no solutions exist in this case, as it becomes evident after realizing that algebraic condition \eref{Sutherland_KPS} is
insufficient of providing \textit{traceless} operators in boundary local spaces, after partially tracing over the auxiliary space $\mathfrak{H}_{a}$.
Accordingly, at this stage we shall simply restrict ourselves to Cholesky form of solutions $\rho_{\infty}=S_{n}S_{n}^{\dagger}$, which we
argue encodes a \textit{minimal} ansatz within a \textit{single}-MPO description for the NESS density operator.
More precisely, as the operator $S_{n}S_{n}^{\dagger}$ can indeed be represented as contraction of ``doubled'' monodromy operator
operating in $\mathfrak{H}_{s}\otimes \mathfrak{H}_{a}$, with $\mathfrak{H}_{a}=\mathfrak{H}_{1}\otimes \mathfrak{H}_{2}$,
we have to demand irreducibility of the auxiliary space as well.

Expanding $\adH(\rho_{\infty})$ by means of Leibniz rule we have
\begin{equation}
\ii \adH(\rho_{\infty})=-\ii S_n(\adH(S_n))^{\dagger}+\ii \adH(S_n)S_n^{\dagger}=
\DD_{\rm L}(SS^{\dagger})+\DD_{\rm R}(SS^{\dagger}).
\label{adH_Leibnitz}
\end{equation}
By employing MPO form \eref{ansatz_S}, in conjunction with \eref{Sutherland_equation}, we obtain
\begin{eqnarray}
\adH(S_n)&=\bra{l}[H,\bb{L}_{1}\cdots \bb{L}_{n}]\ket{r}\nonumber \\
&=\sum_{j=1}^{n-1}\bra{l}\bb{L}_{1}\cdots \bb{L}_{j-1}[h_{j,j+1},\bb{L}_{j}\bb{L}_{j+1}]\bb{L}_{j+2}\cdots \bb{L}_{n}\ket{r}\\
&=\bra{l}\bb{B}_{1}\bb{L}_{2}\cdots \bb{L}_{n}\ket{r}-\bra{l}\bb{L}_{1}\cdots \bb{L}_{n-1}\bb{B}\ket{r}=:S_n^{(\rm L)}-S_n^{(\rm R)},\nonumber
\end{eqnarray}
introducing two boundary \textit{defect-operators} $S^{(L,R)}$. With aid of this result, we rewrite \eref{adH_Leibnitz} in the form
\begin{equation}
\ii \left(S_{n}^{(\rm L)}S_{n}^{\dagger}-S_{n}(S_{n}^{(\rm L)})^{\dagger}\right)+\ii\left(S_{n}(S_{n}^{(\rm R)})^{\dagger}-S_{n}^{(\rm R)}S_{n}^{\dagger}\right)=
\DD_{\rm L}(S_{n}S_{n}^{\dagger})+\DD_{\rm R}(S_{n}S_{n}^{\dagger}).
\label{Lindblad_left_right}
\end{equation}
The key idea now is to use locality of the dissipation and solve \eref{Lindblad_left_right} by imposing a \textit{stronger} condition via identification of two separate
\textit{boundary equations}, written out in fully expanded form as
\begin{eqnarray}
\bra{l}\otimes \bra{l}\left(\ii \mathbb{B}^{(1)}_{1}-\ii \mathbb{B}^{(2)}_{1}-
\DD_{\rm L}(\mathbb{L}_{1})\right)
\left(\prod_{k=2}^{n}\mathbb{L}_{k}\right)\ket{r}\otimes \ket{r}=0,\nonumber \\
\bra{l}\otimes \bra{l}\left(\prod_{k=1}^{n-1}\mathbb{L}_{k}\right)
\left(\ii \mathbb{B}^{(1)}_{n}-\ii \mathbb{B}^{(2)}_{n}+
\DD_{\rm R}(\mathbb{L}_{n})\right)\ket{r}\otimes \ket{r}=0,
% just decoupled equation from above
%\ii \left(S_{n}^{(\rm L)}S_{n}^{\dagger}-S_{n}(S_{n}^{(\rm L)})^{\dagger}\right)=\DD_{\rm L}(S_{n}S_{n}^{\dagger}),\nonumber \\
%\ii\left(S_{n}(S_{n}^{(\rm R)})^{\dagger}-S_{n}^{(\rm R)}S_{n}^{\dagger}\right)=\DD_{\rm R}(S_{n}S_{n}^{\dagger}).
\label{boundary_equations}
\end{eqnarray}
with shorthanded notation $\mathbb{L}_{j}:=\bb{L}_{1,j}\ol{\bb{L}}^{t}_{2,j}$,
$\mathbb{B}^{(1)}_{j}:=\bb{B}_{1,j}\ol{\bb{L}}^{t}_{2,j}$,
$\mathbb{B}^{(2)}_{j}:=\bb{L}_{1,j}\ol{\bb{B}}^{t}_{2,j}$,
%and $\bra{l}\otimes \bra{l}\equiv \bra{l,l}$, $\ket{r}\otimes \ket{r}\equiv \ket{r,r}$.
and superscript $t$ denoting the transposition of the local physical space $\mathfrak{h}_{j}$ and overlined symbol conjugated operators.
In fact, we may factor out the 'free parts', and instead require annihilation of partially-contracted expressions, i.e.
\begin{eqnarray}
% partially contracted equations
\bra{l}\otimes \bra{l}\left(\ii \mathbb{B}^{(1)}_{1}-\ii \mathbb{B}^{(2)}_{1}-\DD_{\rm L}(\mathbb{L}_{1})\right)&=0,\nonumber \\
\left(\ii \mathbb{B}^{(1)}_{n}-\ii \mathbb{B}^{(2)}_{n}+\DD_{\rm R}(\mathbb{L}_{n})\right)\ket{r}\otimes \ket{r}&=0.
% compact notation without reference to two-fold auxiliary space
% \bra{l}\otimes \bra{l}\left(\ii \mathbb{B}^{(1)}_{1}-\ii \mathbb{B}^{(2)}_{1}-
% \DD_{\rm L}(\mathbb{L}_{1})\right)
% \prod_{k=2}^{n}\mathbb{L}_{k}\ket{r}\otimes \ket{r}=0\nonumber \\
% \bra{l}\otimes \bra{l}\prod_{k=1}^{n-1}\mathbb{L}_{k}
% \left(\ii \mathbb{B}^{(1)}_{n}-\ii \mathbb{B}^{(2)}_{n}-
% \DD_{\rm L}(\mathbb{L}_{n})\right)\ket{r}\otimes \ket{r}=0,
% with (a_{1},a_{2}) notation
%\bra{l,l}\left(\ii \mathbb{B}^{(1)}_{a_{1},a_{2},1}-\ii \mathbb{B}^{(2)}_{a_{1},a_{2},1}-
%\boldsymbol{\DD}_{L}(\mathbb{L}_{a_{1},a_{2},1})\right)
%\left(\prod_{k=2}^{n}\mathbb{L}_{a_{1},a_{2},k}\right)\ket{r,r}=0\nonumber \\
%\bra{l,l}\left(\prod_{k=1}^{n-1}\mathbb{L}_{a_{1},a_{2},k}\right)
%\left(\ii \mathbb{B}^{(1)}_{a_{1},a_{2},n}-\ii \mathbb{B}^{(2)}_{a_{1},a_{2},n}-
%\boldsymbol{\DD}_{L}(\mathbb{L}_{a_{1},a_{2},n})\right)\ket{r,r}=0,
\label{boundary_compact}
\end{eqnarray}
Thus, the task has been reduced to find a parametrization of a set of auxiliary matrices $L^{jk}$ and appropriate auxiliary boundary
bra-vector $\bra{l}$ and ket-vector $\ket{r}$ which solve \eref{boundary_compact} for a specific set of dissipative channels $\{A_{\mu}\}$.

%with shorthanded notation $\mathbb{L}_{a_{1},a_{2},j}:=\bb{L}_{a_{1},j}\ol{\bb{L}}^{t_s}_{a_{2},j}$,
%$\mathbb{B}^{(1)}_{a_{1},a_{2},j}:=\bb{B}_{a_{1},j}\ol{\bb{L}}^{t_s}_{a_{2},j}$,
%$\mathbb{B}^{(2)}_{a_{1},a_{2},j}:=\bb{L}_{a_{1},j}\ol{\bb{B}}^{t_s}_{a_{2},j}$,

%For instance, taking the dissipators \eref{Tomaz_dissipators}, yields the solution for the Hamiltonian \eref{Hamiltonian_XXZ} of the form
%\begin{equation}
%L(\epsilon)=\sigma^{0}\otimes \mm{A}(\epsilon)+\sigma^{+}\otimes \bb{A}^{+}(\epsilon)+\sigma^{-}\otimes \bb{A}^{-}(\epsilon),
%\end{equation}
%with the triple of MPO matrices $\{\mm{A}_{\pm},\mm{A}_{0}\}$ whose explicit form can be
%found in [PRL107], contracted with respect to left and right vacuum states, $\bra{L}=\bra{0}$ and $\ket{R}=\ket{0}$.

We have to stress at this stage that it is not a-priori clear whether symmetric dissipators considered in \cite{PRL107}
exhaust \textit{all} possibilities which yield a solution of \eref{boundary_compact} for the model discussed in \Sref{sec:lindblad}.
To this end, one has to incorporate the most general form of a local dissipator in $\mathfrak{h}_{m}$ of the Lindblad form,
\begin{equation}
\DD(\rho)=\sum_{i,j}\sum_{k,l}\GG^{ij}_{kl}\left(E_{m}^{ij}\rho(E_{m}^{kl})^{\dagger}-
\frac{1}{2}\left\{(E_{m}^{kl})^{\dagger}E_{m}^{ij},\rho\right\}\right),
\label{general_dissipator}
\end{equation}
where $\GG=\GG^{\dagger}$ is a \textit{positive} \textit{rate-matrix}, $\GG\geq 0$.

\subsection{Lowest weight representations (Verma modules)}
\label{sec:Verma}
It is now clear that in order to solve the boundary conditions \eref{boundary_compact} we have to be able to freely choose
representation parameters of the Lax and boundary operators, since equations \eref{boundary_compact} may be solvable only by using a
particular representation. Therefore, we have to find a solution of the Sutherland equation \eref{boundary_equations}
(e.g. by using the FRT construction outlined in \Sref{sec:QG})  in terms of abstract objects, i.e. ${\bf L}_j$, which can be
evaluated in generic representation spaces (not necessary finite-dimensional). In other words, in order to guarantee a continuous family of solutions, it is
crucial to allow for (spin) representations going beyond standard finite-dimensional unitary representations. 
For this purpose we shall employ \textit{Verma modules} (see e.g. \cite{Hall03}) -- those are, for generic set of representation parameters (i.e. weight vector),
\textit{infinite-dimensional} irreducible \textit{non-unitary} representations, admitting a lowest weight vector -- with
no associated simply-connected compact group. Therefore, since Verma representations describe our ancillary particles we might, loosely speaking,
proclaim them as a \textit{non-compact spins}.
\paragraph{Example: representation of $\Uqsl{2,\CC}$}
Focusing for the moment on our paradigmatic example exhibiting $\Uqsl{2,\CC}$ symmetry, we define a representation space
$\mathfrak{S}_{s}$, spanned by an infinite basis of state vectors $\{v_{k}\}_{k=0}^{\infty}$, and labeled by a complex (spin)
representation parameter $s\in \CC$, designating the lowest weight, and Casimir invariant $C^{(2)}=s(s+1)$.
A standard realization~\cite{KKM} of such module is a space $\CC[x]$ of polynomials
in variable $x$ with canonical monomial basis $\{v_k=x^{k}=:\ket{k}\}$ and lowest weight $v_{0}=1$. Spin generators are given by
$q$-differential operators (denoting $\partial=\partial/\partial x$)
\begin{equation}
% a-la Dobrev (representation parameter is parametrized as r->2s, where s is SL(2) spin parameter)
S^{z}_{q}=x\partial-s,\quad S^{+}_{q}=x[2s-x\partial]_q,\quad S_{q}^{-}=x^{-1}[x\partial]_{q}.
% a-la Derkachov
%\bb{S}^{z}_{q}=x\partial+s,\quad \bb{S}^{+}_{q}=x[x\partial+2s]_q,\quad \bb{S}_{q}^{-}=-x^{-1}[x\partial]_{q}.
\end{equation}
% C_q = S_{q}^{+}S_{q}^{-}+[S^{z}]_q[S^{z}-1]_q
In the undeformed case $q\rightarrow 1$ we recover $\mathfrak{sl}(2,\CC)$ generators as simple first-order differential operators
\begin{equation}
% a-la Dobrev
S^{z}=x\partial-s,\quad S^{+}=-x^{2}\partial+2sx,\quad S^{-}=\partial,
% a-la Derkachov
%\bb{S}^{z}=x\partial+s,\quad \bb{S}^{+}=x^{2}\partial+2sx,\quad \bb{S}^{-}=-\partial,
\end{equation}
and quadratic central element (Casimir operator) $C=(S^{z})^{2}-S^{z}+S^{+}S^{-}$.
% we drop global transformations
% which generate \textit{global transformations} (with parameter $\xi\in \CC$) on polynomials $f(x)$,
% \begin{eqnarray}
% \xi^{\bb{S}^{z}}f(x)&=\xi^{s}f(\xi x),\nonumber \\
% e^{\xi \bb{S}^{-}}f(x)&=f(x-\xi),\nonumber \\
% e^{\xi \bb{S}^{+}}f(x)&=(1-\xi x)^{-2s}f\left(\frac{x}{1-\xi x}\right),
% \end{eqnarray}
% generalizing translations, dilatations and special conformal transformation, constituting a group of M\"{o}bius transformations.
The basis vectors of $\mathfrak{S}_{s}$ are constructed by means of the generating function -- a coherent state vector --
\begin{equation}
% a-la Dobrev
e^{\xi S^{+}}v_{0}=\sum_{k=0}^{\infty}\frac{\xi^{k}}{k!}(S^{+})^{k}v_{0}=\sum_{k=0}^{\infty}\frac{\xi^{k}}{k!}(2s)_{k}v_{k},
%(2s)_{k}=\frac{\Gamma(s+1)}{\Gamma(s-k+1)}
% a-la Derkachov
%e^{\xi \bb{S}^{+}}v_{0}=\sum_{k=0}^{\infty}\frac{\xi^{k}}{k!}(\bb{S}^{+})^{k}v_{0}=\sum_{k=0}^{\infty}\frac{\xi^{k}}{k!}(2s)_{k}v_{k},\qquad
%(2s)_{k}=\frac{\Gamma(2s+k)}{\Gamma(2s)}
\end{equation}
with falling factorial $(2s)_{k}=(2s)(2s-1)\cdots (2s-k+1)$, producing an infinite sequence of states for generic
$s\neq \frac{\ell}{2}$ ($\ell=\ZZ_{+}\equiv\{0,1,2,\ldots\}$),
whereas in the opposite case ($s=\frac{\ell}{2}$) there exists a highest-weight state as well and a module reduces to
\textit{finite} $(\ell+1)$-dimensional \textit{invariant subspaces} $\mathfrak{S}_{\ell}\subset \mathfrak{S}_{s}$ spanned by basis elements $\{x^{k}\}_{k=0}^{n}$.
Those representations are equivalent to unitary representation of $\mathfrak{su}(2,\CC)$.
For instance, $s=\half$ corresponds to the $\mathfrak{S}_{f}:=\mathfrak{S}_{\half}\cong \CC^{2}$, being the \textit{fundamental spin representation} with
standard (Pauli) generators $S^{+}=\sigma^{+}, S^{-}=\sigma^{-}$ and $S^{z}=\half \sigma^{z}$.
Another type of finite dimensional representations, with no classical correspondence, occurs after algebra quantization for $q$ being a
root of unity $q=e^{\ii \gamma}$, $\gamma=\frac{\pi l}{m}$ ($l,m\in \mathbb{N}$), referred to as cyclic representations
\footnote{In cases when representation becomes reducible we implicitly consider restriction to invariant subspace only.}.

The dual vector space (bra-vectors) is defined via basis $\{\bra{l}\}$, such that bi-orthogonality relation $\braket{l}{k}=\delta_{l,k}$ holds.

\paragraph{Product representations.}
The $R$-matrix associated to the algebra of $\{T^{i}_{j}(s)\}$ operates in a product space $\mathfrak{S}_{s_{1}}\otimes \mathfrak{S}_{s_{2}}$.
A generic product representation ($s_{1},s_{2}\not\in \half \mathbb{Z}_{+}$) is a space of polynomials in two variables $\CC[x,y]$ and is
simply-decomposable as a direct sum of infinite-dimensional lowest weight spaces,
\begin{equation}
\mathfrak{S}_{s_{1}}\otimes \mathfrak{S}_{s_{2}}=\bigoplus_{\nu=0}^{\infty}\mathfrak{S}_{s_{1}+s_{2}-\nu},
% a-la Derkachov
%\mathfrak{S}_{s_{1}}\otimes \mathfrak{S}_{s_{2}}=\bigoplus_{\nu=0}^{\infty}\mathfrak{S}_{s_{1}+s_{2}+\nu}.
\label{module_decomposition}
\end{equation}
labeled by index $\nu \in \mathbb{Z}_{+}$. For spin parameters being \textit{both} positive half-integers,
$s_{1},s_{2}\in \half \mathbb{Z}_{+}$, we have a
finite number of factors in decomposition \eref{module_decomposition}, labeled as $\nu=0,1,\ldots 2\min\{s_{1},s_{2}\}$.
The lowest-weight vectors $\psi^{0}_{\nu}(x,y)=(x-y)^{\nu}$ are annihilated upon action of the global
lowering generator $S^{-}=S^{-}_{1}+S^{-}_{2}$,
\begin{eqnarray}
S^{-}(s_{1},s_{2})\psi^{0}_{\nu}(x,y)&=0,\nonumber \\
S^{z}(s_{1},s_{2})\psi^{0}_{\nu}(x,y)&=(s_{1}+s_{2}-\nu)\psi^{0}_{\nu}(x,y),
\end{eqnarray}
whereas the infinite tower of states $\{\psi_{\nu}^{m}(x,y)\}$ spanning $\mathfrak{S}_{s_{1}+s_{2}-\nu}$ are given by the
action of the raising operator
$S^{+}=S^{+}_{1}(s_{1},s_{2})+S^{+}_{2}(s_{1},s_{2})$,
\begin{equation}
\psi^{m}_{\nu}(x,y)=(S^{+})^{m}\psi^{0}_{\nu}(x,y).
\end{equation}

\paragraph{$\Uqsl{N,\CC}$ representations.}
For higher-dimensional quantum spaces we approach \`{a}-la Dobrev~\cite{Dobrev} and, similarly as in $N=2$ case, consider realization of algebra
generators in terms of linear differential operators -- now operating in a space of polynomials in
$N(N-1)/2$ commuting variables $x_{i}^{k}$, $1\leq i\leq k-1,2\leq k\leq N$ -- with associated number operators $N_{i}^{k}$, defined such
that $N_{i}^{k}x_{j}^{l}=\delta_{ij}\delta_{kl}x_{j}^{l}$, and corresponding $q$-differential operators $D_{i}^{k}=(x_{i}^{k})^{-1}[N_{i}^{k}]_q$.
We chose to work with a generic   $N$-dimensional representation parameter (weight) vector $\underline{r}=(r_{0},r_{1},\ldots,r_{N-1})$, $r_{k}\in \CC$,
chosen with convention that representation is irreducible \textit{iff} all $r_{i}$ are non-negative integers.
In this picture, a value of the central element $\lambda$ pertains to the combination $\lambda=\sum_{i=0}^{N-1}(N-i)r_{i}$.
For instance, in the $N=2$ case above, we have a single spin representation parameter $r_{1}=2s$ and a spectral parameter $\lambda=2r_{0}+r_{1}$.

For further details and canonical construction of $\Uqsl{n,\CC}$ Verma modules we refer the reader to the reference\cite{Dobrev}.

In the following we shall use the approach outlined above to find steady states of models with $\Uqsl{2,\CC}$ deformed symmetry for spin-$1/2$
Hamiltonians, as well as $SU(N)$-symmetric hopping Hamiltonians with local physical space of dimension $N$.
\subsection{Solution with $\Uqsl{2,\CC}$ deformed symmetry}
% note how global H with open boundary condition is not Uqsl(2)-invariant, but only U(1), due to absence of surface terms

The Baxterized $R$-matrix explicitly reads
\begin{equation}
R^{q}_{12}(\lambda)=(q-q^{-1})\pmatrix{
[-\ii \lambda+1]_q & & & \cr
& [-\ii \lambda]_q & e^{\gamma \lambda} & \cr
& e^{-\gamma \lambda} & [-\ii \lambda]_q & \cr
& & & [-\ii \lambda+1]_q
},
\label{R_baxterized}
\end{equation}
and becomes, after taking the derivative with respect to $\lambda$ at $\lambda=0$, and subsequently left-multiplying by $P_{12}$,
\begin{equation}
\partial_{\lambda}\PR_{12}^{q}(\lambda)|_{\lambda=0}=\gamma
\pmatrix{
q+q^{-1} & & & \cr
& -(q-q^{-1}) & 2 & \cr
& 2 & q-q^{-1} & \cr
& & & q+q^{-1}
}.
\end{equation}
With the definition of anisotropy parameter $\gamma$, $2\cos{(\gamma)}=q+q^{-1}$ and $2\ii \sin{(\gamma)}=q-q^{-1}$,
we express this result in terms of Pauli spin variables, and obtain the $2$-body $\mathcal{U}_{q}(\mathfrak{su}(2))$-covariant
interaction
\begin{equation}
h_{12}=\gamma\left[2\sigma^{+}_{1}\sigma^{-}_{2}+2\sigma^{-}_{1}\sigma^{+}_{2}+
\cos{\gamma}(\sigma^{0}_{1}\sigma^{0}_{2}+\sigma^{z}_{1}\sigma^{z}_{2})-
\ii \sin{\gamma}(\sigma^{z}_{1}\sigma^{0}_{2}-\sigma^{0}_{1}\sigma^{z}_{2})\right].
\label{nonhermitian_interaction}
\end{equation}
Plugging the interaction \eref{nonhermitian_interaction} into \eref{Sutherland_equation}, in conjunction with redefinitions
$\bb{L}_{j}(\lambda)\rightarrow (2\ii \gamma)^{-1}\bb{L}_{j}(\lambda)$ and
$R^{q}_{12}(\lambda)\rightarrow \gamma^{-1}R^{q}_{12}(\lambda)$,
with $\PR_{12}(0)=2\ii\sinc{\gamma}\mathds{1}_{12}$ and employing definition $\sinc{x}:=\sin{(x)}/x$, we finally obtain the
Sutherland equation \eref{Sutherland_equation}, with the Lax operator
\begin{equation}
%\bb{L}(\lambda)=\frac{1}{\gamma}\pmatrix{
%\sin{(\varphi+\gamma S^{z})} & e^{\gamma \lambda}\sin{(\gamma)}S^{-} \cr
%e^{-\gamma \lambda}\sin{(\gamma)}S^{+} & \sin{(\varphi -\gamma S^{z})}
%},\quad \varphi:=-\ii \gamma \lambda
\bb{L}_{j}(\lambda)=\sinc{\gamma}\pmatrix{
[-\ii \lambda+S^{z})]_q & e^{\gamma \lambda}S^{-} \cr
e^{-\gamma \lambda}S^{+} & [-\ii \lambda-S^{z}]_q
}_{j}
\label{XXZ_Lax}
\end{equation}
and the boundary operator $\bb{B}(\lambda)$ which admits the form
\begin{eqnarray}
\bb{B}_{j}(\lambda)&:=-\PR(0)\partial_{\lambda}\bb{L}_{j}(\lambda)=-2\ii\sinc{\gamma}\partial_{\lambda}\bb{L}_{j}(\lambda)\nonumber \\
&=-2\sinc{\gamma}\pmatrix{
\cos{[\gamma(-\ii \lambda+S^{z})]} & \ii\sin{(\gamma)}e^{\gamma \lambda}S^{-} \cr
-\ii\sin{(\gamma)}e^{-\gamma \lambda}S^{+} & \cos{[\gamma(-\ii \lambda+S^{z})]}
}.
\end{eqnarray}
The normalization has been chosen such that $q\rightarrow 1$ ($\gamma\rightarrow 0$) undeformed limit recovers the $\mathfrak{sl}_{2}$-covariant
Lax operator (writing $u:=-\ii \lambda$)
\begin{eqnarray}
\lim_{\gamma\rightarrow 0}\bb{L}_{j}(\lambda)&=
\pmatrix{
u+S^{z} & S^{-} \cr
S^{+} & u-S^{z}
}=
u\cdot \sigma^{0}_{j}\otimes \cdot\mathds{1}+\vec{\sigma}_{j} \otimes \vec{S},\nonumber \\
\lim_{\gamma\rightarrow 0}\bb{B}_{j}(\lambda)&=-2\cdot\sigma_{j}^{0}\otimes \mathds{1},
\label{XXX_Lax}
\end{eqnarray}
compliant with the $SU(2)$-invariant interaction, $\lim_{\gamma \rightarrow 0}(\gamma^{-1}h_{12})=2P_{12}$.

It is important to stress that $h_{j,j+1}$ is \textit{not hermitian}, therefore, not in the right form for application we have in mind.
Nevertheless, we show how the problem can easily be circumvented at the level of the Sutherland equation, resting on the fact that
the hermicity is violated only by an anti-hermitian \textit{surface term}, namely the anisotropic Heisenberg interaction can be expressed as
\begin{equation}
h^{XXZ}_{j,j+1}=h_{j,j+1}-\ii\sinc{\gamma}\left(\sigma^{z}_{j}\sigma^{0}_{j+1}-\sigma^{0}_{j}\sigma^{z}_{j+1}\right).
\end{equation}
Consequently, we expand the commutator on the right side of \eref{Sutherland_equation},
\begin{equation}
[h_{j,j+1},\bb{L}_{j}\bb{L}_{j+1}]=[h^{XXZ}_{j,j+1},\bb{L}_{j}\bb{L}_{j+1}]
+\ii\left([b,\bb{L}]_{j}\bb{L}_{j+1}-\bb{L}_{j}[b,\bb{L}]_{j+1}\right),
\end{equation}
by means of a local operator $b_{j}=-\sinc{\gamma}\sigma^{z}_{j}$, and absorb the surface term
$\bb{B}_{j}^{\partial}(\lambda):=[b,\bb{L}_{j}(\lambda)]$ into \textit{redefinition} of the boundary operator,
\begin{eqnarray}
\bb{B}^{XXZ}_{j}(\lambda)&=\bb{B}_{j}(\lambda)+\bb{B}_{j}^{\partial}(\lambda)\nonumber \\
&=-2\sinc{\gamma}
\pmatrix{
\cos{(\gamma(-\ii \lambda+S^{z}))} & 0 \cr
0 & \cos{(\gamma(-\ii \lambda+S^{z}))}
}_{j}.
%\pmatrix{
%\cos{(\gamma(-\ii \lambda+\bb{S}^{z}))} & 0 \cr
%0 & \cos{(\gamma(-\ii \lambda-\bb{S}^{z}))}
%}_{j}.
\end{eqnarray}
%introducing shorthanded $q$-operator symbol $\lshad x\rshad_q:=\cos{(\gamma x)}$.

The $L$-operator remains in principle intact in this case, however, for convenience we may remove exponential scalar factors $e^{\pm \gamma \lambda}$
from off-diagonal elements as they are merely expressing $\lambda$-dependent transformation defining a spin algebra automorphism, i.e. we employ
the Lax operator in the form of
\begin{equation}
\bb{L}^{XXZ}_{j}(\lambda)=\sinc{\gamma}\pmatrix{
[-\ii \lambda+S^{z}]_{q} & S^{-} \cr
S^{+} & [-\ii \lambda-S^{z}]_{q}.
}_{j}
\end{equation}
An alternative (more formal) and equivalent procedure to produce hermitian interaction involves \textit{twisting} a Hopf algebra, generating new pair of
$R-$ and $L$-operators satisfying local Yang-Baxter equation \eref{local_YBE}, which is briefly outlined in the Appendix \ref{App:twist}.

%We remark, that performing Baxterization of constant solutions of YBE is of crucial importance for passing to required
%operator-divergence relation.

%\begin{equation}
%[h^{XXZ},L(\lambda)\otimes_{s}L(\lambda)]=
%B(\lambda)\otimes_{s}L(\lambda)-L(\lambda)\otimes_{s}B(\lambda).
% -2\left(\frac{\sin{\gamma}}{\gamma}\right)
%\label{XXZ_Sutherland}
%\end{equation}

% for for Lie symmetry sl(n) one obtains simple the gl(n) defining relation for element of L.

For the sake of clarity, we shall consistently omit the $XXZ$ superscript on the operators $\bb{L}^{XXZ}_{j}(\lambda),\bb{B}^{XXZ}_{j}(\lambda)$
in the subsequent discussion. We hope this choice will not lead to confusion with $\Uqsl{2,\CC}$-symmetric objects associated with the
non-hermitian interaction $h_{12}$ from \eref{nonhermitian_interaction}. Thus, we proceed by evaluating the \textit{non-vanishing} action of the
Lax matrix on the boundary vacuum states. We have (omitting momentarily dependence on representation parameters)
\begin{eqnarray}
% Dobrev with r1=2s
\label{contraction_L}
L^{11}\ket{0}&=[-\ii \lambda-s]_q\ket{0},& \bra{0}L^{11}=[-\ii \lambda-s]_q\bra{0},\nonumber \\
L^{22}\ket{0}&=[-\ii \lambda+s]_q\ket{0},& \bra{0}L^{22}=[-\ii \lambda+s]_q\bra{0},\nonumber \\
L^{21}\ket{0}&=0,& \bra{0}L^{21}=\sinc{\gamma}\bra{0},\nonumber \\
L^{12}\ket{0}&=\sinc{\gamma}[2s]_{q}\ket{1},& \bra{0}L^{12}=0,
% Derkachov
% \bb{L}^{00}\ket{0}&=[-\ii \lambda+s]_q\ket{0},\quad \bra{0}\bb{L}^{00}=[-\ii \lambda+s]_q\bra{0},\nonumber \\
% \bb{L}^{11}\ket{0}&=[-\ii \lambda-s]_q\ket{0},\quad \bra{0}\bb{L}^{11}=[-\ii \lambda-s]_q\bra{0},\nonumber \\
% \bb{L}^{01}\ket{0}&=0,\quad \bra{0}\bb{L}^{01}=-\sinc{\gamma}\bra{0},\nonumber \\
% \bb{L}^{10}\ket{0}&=\sinc{\gamma}[2s]_{q}\ket{1},\quad \bra{0}\bb{L}^{10}=0,
\end{eqnarray}
whereas the boundary matrix gives
\begin{eqnarray}
% Dobrev with r1=2s
\label{contraction_B}
B^{11}\ket{0}=-2\sinc{\gamma}\cos{[\gamma(-\ii \lambda-s)]},\nonumber \\
B^{22}\ket{0}=-2\sinc{\gamma}\cos{[\gamma(-\ii \lambda+s)]},
% Derkachov 
%\bb{B}^{00}\ket{0}=-2\sinc{\gamma}\cos{[\gamma(-\ii \lambda+s)]},\nonumber \\
%\bb{B}^{11}\ket{0}=-2\sinc{\gamma}\cos{[\gamma(-\ii \lambda-s)]},
\end{eqnarray}
operating equally on the bra-vector $\bra{0}$.
% \quad \bra{0}\bb{B}^{00}=-2\sinc{\gamma}\cos{[\gamma(-\ii \lambda+s)]}\bra{0}\nonumber \\
%\quad \bra{0}\bb{B}^{11}=-2\sinc{\gamma}\cos{[\gamma(-\ii \lambda+s)]}\bra{0}.

Evaluating boundary equations \eref{boundary_compact} with general form of the left/right dissipators $\GG_{\rm L}$ and $\GG_{\rm R}$ from
\eref{general_dissipator}, and projecting result onto (Weyl) basis elements in the boundary physical spaces,
i.e. $E^{\alpha\beta}_{1}$ on the left and $E^{\alpha\beta}_{n}$ on the right,
results in the system of \textit{operator-valued} equations for each $\alpha,\beta=1,2$.
Finding a solution to the problem admitting $\Uqsl{N,\CC}$ symmetry thus boils down, accounting hermicity of $\rho_{\infty}$,
to simultaneously satisfy $N(N+1)$ (possibly linearly dependent) equations, for general spin-$s$ Lax operator.

In the case of fixed boundary vectors ($\ket{r}=\ket{0}$ and $\bra{l}=\bra{0}$) considered in \eref{contraction_L}, \eref{contraction_B} the
incoherent maximal driving, implemented via left channel $A_{1}=\sqrt{\Gamma}\sigma^{+}_{1}$ and right channel
$A_{n}=\sqrt{\Gamma}\sigma^{-}_{n}$ is the only dissipator yielding a solution, given by the spin
parameter $s$ related to the coupling via
\begin{equation}
% Dobrev
\label{result_fundamental}
\Gamma=4\sin{(\gamma)}\coth{(\gamma \,{\rm Im}(s))}=4\ii\frac{\cos(\ii\gamma \,{\rm Im}(s))}{[\ii\,{\rm Im}(s)]_q},
% Derkachov 
%\Gamma=-4\sin{\gamma}\coth{(\gamma \sigma)}=(-4\ii)\frac{\cos(\gamma \ii \sigma)}{[\ii \sigma]_q},
\end{equation}
and $\Re(s)=0$, in accordance with solutions~\cite{PRL107,KPS}. The isotropic solution reads $s=4\ii \Gamma^{-1}$.

% add dephasing channels

\subsection{Integrable chains with $SU(N)$ global symmetry}

Presented formalism can be immediately facilitated to address higher-spin isotropic models and their $q$-deformed counterparts.
Let us initially treat the isotropic models, i.e. chains exhibiting global $SU(N)$ symmetry (e.g. higher spin generalization of
the $S=1$ Lai-Sutherland model), $N=2s+1$ ($N=2,3,\ldots$), with local quantum (spin) space
$\mathfrak{h}_{j}\cong \mathfrak{S}_{s}\cong \CC^{N}$.
In this case we avoid using spin variables $S_{j}^{k}$ ($k=x,y,z$) in $\mathfrak{h}_{j}$, since the interaction
can be expressed via permutation operator over two adjacent quantum spaces,
\footnote{Those are not to be confused with higher-spin $SU(2)$-invariant models, which are expressed in terms of polynomials
in quadratic Casimir operator of $\mathfrak{sl}(2,\CC)$ (see e.g.~\cite{Faddeev2,Bytsko}) and are non-fundamental.}
\begin{eqnarray}
H=\sum_{j=1}^{n-1}P_{j,j+1}=\sum_{j=1}^{n-1}\sum_{k,l=1}^{N}E^{kl}_{j}E^{lk}_{j+1}.
\label{H_spin_n}
\end{eqnarray}

% comment that in principle we apply general theory, but for SU(n) this is an overhead since we only need to solve the Sutherland
% for q-deformed counterparts however it is not trivial, but we do not take that pathway because of problem with hermicity

\paragraph{$\mathcal{U}_{q}(\mathfrak{gl}(N,\CC))$-invariant Lax operator.}
We could in principle proceed along the lines of preceding discussion, however since we address the isotropic models,
i.e. interactions described by non-deformed (Lie) symmetries, we rather employ an instructive shortcut to the solution of the Sutherland equation,
entirely avoiding an explicit use of algebraic objects from FRT construction. 

We start by proposing the following expansions for \textit{constant} operators $\bb{L}$ and $\bb{B}$,
\begin{equation}
\bb{L}_{m}=\sum_{i,j=1}^{N}E_{m}^{ij}\otimes L^{ji},\qquad
\bb{B}_{m}=\sum_{i,j=1}^{N}E_{m}^{ij}\otimes B^{ji}.
\end{equation}
After considering interaction $h_{j,j+1}=P_{j,j+1}$ we derive
\begin{eqnarray}
[P_{m,m+1},\bb{L}_{m}\bb{L}_{m+1}]&=\sum_{i,j=1}^{N}\sum_{k,l=1}^{N}E_{m}^{ij}E_{m+1}^{kl}\otimes [L^{jk},L^{li}],\nonumber \\
\bb{B}_{m}\bb{L}_{m+1}-\bb{L}_{m}\bb{B}_{m+1}&=\sum_{i,j=1}^{N}\sum_{k,l=1}^{N}E_{m}^{ij}
E_{m+1}^{kl}\otimes(B^{ji}L^{lk}-L^{ji}B^{lk}),
\end{eqnarray}
which gives rise to quadratic algebraic relations for the matrix elements
\begin{equation}
[L^{jk},L^{li}]=B^{ji}L^{lk}-L^{ji}B^{lk},
\end{equation}
obviously reducing upon restricting $\bb{B}=-\mathds{1}$, i.e. $B_{ij}=\delta_{ij}\cdot (-\mathds{1})$,
to the \textit{defining relations} of $\mathfrak{gl}(N,\CC)$. Moreover, the solution can be
readily extended by including a \textit{central element} (of the form $u\cdot \mathds{1}$, $u\in \CC$) to the operator $\bb{L}$,
enabling us to work with $\mathfrak{gl}(N,\CC)$-covariant Lax matrix reading
\begin{equation}
\bb{L}_{j}(u)=u\cdot \mathds{1}_{n}\otimes {\mathds{1}}+\sum_{k,l=1}^{N}E_{j}^{kl}\otimes L^{lk}.
\label{Lax_sln}
\end{equation}
This result is of course of no surprise since it is implied by the underlying YBE, with fundamental $\mathfrak{gl}(N,\CC)$ $R$-matrix of the form
\begin{equation}
R_{12}(u)=u\cdot \mathds{1}_{12}+\sum_{i,j=1}^{N}E_{1}^{ij}E_{2}^{ji},
\end{equation}
which agrees (up to additive constant) with previously encountered $R$-matrix from the FRT construction \eref{R_glN}.
%In the appendix we show this results arises from a general YBE \eref{general_YBE} by evaluating two space in the fundamental spaces $\mathfrak{S}_{f}$ ($s_{1}=s_{2}=f$) and
%the third space, determining the element of the Lax operator \eref{Lax_sln} in arbitrary representation $\mathfrak{S}_{s_{3}}$.
Notice also that $N=2$ case is consistent with previously obtained result \eref{result_fundamental} upon identifying $h^{XXZ}_{12}=2P_{12}$,
for $P_{12}$ in $\mathfrak{S}_{\half}\otimes \mathfrak{S}_{\half}\cong \CC^{4}$.

\paragraph{Solution of the boundary equations.}
Having facilitated the Lax matrix \eref{Lax_sln}, we can proceed as discussed above and seek for a solution of the boundary equations \eref{boundary_compact}.
%In this section we shall consider a model of multicomponent interacting quantum gas with local quantum space $\mathfrak{h}_{j}\cong \mathbb{C}^{N}$,
%with fully isotropic ($SU(n)$ symmetric) interaction, i.e. a hopping process preserving number of particles of each species, known as
%the \textit{Lai-Sutherland model},
%\begin{eqnarray}
%H=\sum_{j=1}^{n-1}P_{j,j+1}=\sum_{j=1}^{n-1}\sum_{k,l=1}^{N}E^{kl}_{j}E^{lk}_{j+1}.
%\label{H_spin_n}
%\end{eqnarray}
Since a general problem to characterize all integrable steady states is hard even in the simplest case $N=2$,
we shall restrict ourselves in the forthcoming discussion only to {\it maximal incoherent driving} situation,
determined by two sets of Lindblad operators acting at the ends of the chain,
\begin{equation}
\label{A_spin_n}
A_1^{(k)}=\sqrt{\Gamma_{\rm 1}} E^{kN}_1,\quad A_n^{(k)}=\sqrt{\Gamma_{n}}E^{Nk}_n,\quad k=1,2,3,\ldots N-1.
\end{equation}
The Hamiltonian (\ref{H_spin_n}) and the Lindblad operators (\ref{A_spin_n}) generate, by multiplication and addition,
the entire algebra of operators, which in turn implies \textit{uniqueness} of the steady state density operator $\rho_{\infty}$~\cite{Eva77}.
As in the $N=2$ case, we expect that left and  right boundary contraction vectors are in fact  left and right auxiliary vacua which,
in language of Dobrev's representation for the generators $L^{ij}$ \cite{Dobrev}, satisfy
\begin{equation}
\label{vac_spin_n}
\langle 0| x_i^k=0,\quad \partial_{x^k_i}|0 \rangle=0.
\end{equation}
With this choice of vacua and Lindblad operators one can see that the compatibility conditions (\ref{boundary_compact}) further restrict
the weight vector, imposing $r_1=r_2\ldots=r_{N-2}=0$, hence leaving $r_0$ and $r_{N-1}$ as the only non-vanishing representation parameters. 
This further simplifies the Lax matrix
{\small \begin{eqnarray}
\!\!\!\!\!\!\!\!\!\!\!\!\!\!\!\!{\bf L}_{m}&=\sum_{j=1}^{N-1}\left(E_{m}^{jj} (x_j^N\partial_{x_j^N}+r_0) + E_{m}^{jN}\partial_{x_j^N}+
E_{m}^{Nj}x_j^N(-x_j^N\partial_{x_j^N} + r_{N-1})+E_{m}^{NN}x_j^N\partial_{x_j^N} \right)\nonumber \\
\!\!\!\!\!\!\!\!\!\!\!\!\!\!\!\!&~~~+\sum_{j\neq k=1}^{N-1}E_{m}^{kj} x_j^N\partial_{x_k^N}+E_{m}^{NN}(r_0+r_{N-1}).
\label{L_spin_n}
\end{eqnarray}} 
By using properties (\ref{vac_spin_n}) we can evaluate matrices appearing in boundary equations (\ref{boundary_compact}) explicitly.
On the left side we obtain
{\small \begin{eqnarray}
\nonumber 
\bra{0,0}\mathbb{B}^{(1)}_{1}&=-2\sum_{j=1}^{N-1} \left(\bar{r}_0 E_{1}^{jj} \bra{0,0}+E_{1}^{Nj}\bra{0,1_j}\right)+
(\bar{r}_0+\bar{r}_{N-1})E_{1}^{NN}\bra{0,0}, \\ \nonumber
\bra{0,0}\mathbb{B}^{(2)}_{1}&=-2\sum_{j=1}^{N-1}  \left(r_0 E_{1}^{jj} \bra{0,0}+E_{1}^{jN}\bra{1_j,0}\right)+(r_0+r_{N-1})E_{1}^{NN}\bra{0,0},\\ \nonumber
\bra{0,0}\DD_{\rm L}(\mathbb{L}_{1})&=-(N-1)\Gamma^{\rm L}|r_0+r_{N-1}|^2E_{1}^{NN}\bra{0,0}+ \Gamma^{\rm L}
\sum_{j=1}^{N-1}\left(|r_0+r_{N-1}|^2E_{1}^{jj}\bra{0,0}\right.\\ \nonumber
	&~~~-\left. \half(N-1)\left((r_0+r_{N-1})E_{1}^{Nj}\bra{0,1_j}-(\bar{r}_0+\bar{r}_{N-1})E_{1}^{jN}\bra{1_j,0}\right)\right),
\end{eqnarray}}
and similarly on the right,
\begin{eqnarray}\nonumber
\mathbb{B}^{(1)}_{n}\ket{0,0}&=-2\sum_{j=1}^{N-1}   (\ket{0,0} \bar{r}_0 E_{n}^{jj} +\bar{r}_{N-1}\ket{0,1_j} E_{n}^{jN})+\ket{0,0}(\bar{r}_0+\bar{r}_{N-1})E_{n}^{NN}, \\ \nonumber
\mathbb{B}^{(2)}_{n}\ket{0,0}&=-2\sum_{j=1}^{N-1}   (\ket{0,0}r_0 E_{n}^{jj} +r_{N-1}\ket{1_j,0} E_{n}^{Nj})+\ket{0,0}(r_0+r_{N-1})E_{n}^{NN},\\ \nonumber
\DD_{\rm L}(\mathbb{L}_{n})\ket{0,0}&=\ket{0,0}(N-1)^{3}|r_{0}|^{2}\Gamma^{\rm R}E_{n}^{NN}-\Gamma^{\rm R}\sum_{j=1}^{n-1}\bigg((N-1)^{2}|r_{0}|^{2}\ket{0,0}E_{n}^{jj} \\ \nonumber
&~~~~  +\half(N-1)^{2}\left(\ket{1_j,0}\bar{r}_0r_{N-1}E_{n}^{Nj} +\ket{0,1_j}r_0\bar{r}_{N-1}E_{n}^{jN}\right) \bigg).
\end{eqnarray}
We introduced a braket notation to denote vectors in a doubled auxiliary space $\mathfrak{H}_{1}\otimes \mathfrak{H}_{2}$, namely we identify
$\ket{0,0}\equiv 1,$ $\ket{1_k,0}\equiv x_k^N$, and $\ket{0,1_k}\equiv y_k^N$ and
similarly for the bra-vectors. In accordance with \eref{boundary_compact} we arrive at the result
\begin{equation}
r_0=-\frac{4\ii}{(N-1)^2 \Gamma},\qquad r_{N-1}=-N r_0,
\end{equation}
together with the coupling strengths of left and right dissipators $\Gamma^{\rm L}=\Gamma$ and $\Gamma^{\rm R}=(N-1)^2\Gamma$, respectively.
It is worth noticing that the obtained solution is a particular generalization of the steady state of the maximally-driven isotropic
spin-$1/2$ Heisenberg model, as they only involve a single non-zero $\Uqsl{N,\CC}$ representation parameter $r_{N-1}$ (being equivalent
to the \textit{spin} parameter in $N=2$ situation) and vanishing spectral parameter $\lambda=Nr_{0}+r_{N-1}=0$,
consequently leading to the same asymptotic particle density profiles, as provided in \Sref{sec:observables}.
%However, in contrast to the $n=2$ case we expect  due to a richer structure of the ansatz the appearance of new solutions
%that are not reminiscent of the maximal driving solution.
%This difficult task of full classification of solutions with $SU(N)$ bulk symmetry, however, remains an open problem.
Unlike in the $N=2$ case, it appears difficult already for $N=3$ to solve \eref{boundary_compact} in its full generality,
especially ultimately verifying if solutions pertain to \textit{positive} rate matrices $\GG_{\rm L,R}$, thus representing realistic noise process.

% comments on drawbacks...
% Analogously, by employing $\mathfrak{gl}(n,\CC)$ Verma representation, one might treat $SU(n)$-invariant models
% for $n>4$ as well. At the moment however, we are only able to demonstrate existence of solution which generalize maximal incoherent driving
% case from $n=3$.
% ... where all but one representation parameter attain non-zero value, thereby significantly simplifying structure of solutions.
% Full classification of solutions with $SU(n)$ bulk symmetry thus remains an open problem.

On the other hand, there is an issue with $q$-deformation when $N>2$; it turns out that hermicity violation of the interaction
is not of surface type, hence disallowing for a simple amendment of the Sutherland equation (or alternatively applying a twist on a
level of Hopf algebra), and consequently rendering the corresponding higher anisotropic models, i.e. the $N$-component
Perk-Schultz models,
\begin{equation}
H^{q}_{\rm PS}=\sum_{j=1}^{n-1}\sum_{\alpha=1}^{N-1}\sum_{\beta=\alpha+1}^{N}
E^{\alpha\beta}_{j}E^{\beta\alpha}_{j+1}+E^{\beta\alpha}_{j}E^{\alpha\beta}_{j+1}-
qE^{\alpha\alpha}_{j}E^{\beta\beta}_{j+1}-q^{-1}E^{\beta\beta}_{j}E^{\alpha\alpha}_{j+1}
\end{equation}
not applicable for description of boundary-driven open quantum systems.
%\footnote{The non-hermitian bulk terms can perhaps be corrected
%by adding dissipation in the bulk, however, the extension of the Sutherland equation to dissipative processes is not the goal of the paper.}

\paragraph{Non-fundamental models.}
We stress that presented technique for constructing solutions from the bulk divergence algebraic condition (Sutherland equation) from
universal $R$-matrix is only possible when the so-called \textit{fundamental models} are addressed, namely the integrable models with
Lax operators whose physical and auxiliary spaces are isomorphic -- the reason being simply that auxiliary spin label from the
Lax operator is reserved for a generic non-compact spin (needed to be tuned with the dissipation parameters), whereas the remaining two
spin labels from the YBE have to be the same in order to discuss homogeneous models. The latter two spin labels define the auxiliary indices with
respect to FRT construction which are eventually, as we have seen, interpreted as local physical indices of two adjacent quantum spaces.
For non-fundamental models, on the other hand, the interaction cannot simply be  deduced from regularity property through derivative of
the $R$-matrix, but instead a pathway via analytic properties of a monodromy matrix needs to be utilized (see e.g. \cite{KunduRev}).

\section{Expectation values of observables}
\label{sec:observables}
An important aspect of analytic exact solutions for the NESS is to be able to efficiently compute expectation values of local physical
observables. For this purpose we may define generic local vertex operators $\mathbb{X}_{N}$ as elements over
$\mathfrak{S}_{s}\otimes \mathfrak{S}_{\overline{s}}$ -- for a system of size $n$ with local space $\mathfrak{h}\cong \CC^{N}$ -- associated
to a local physical operator $X_{j}$ supported on a contiguous sub-lattice consisting of say $k$ physical sites between positions
$j$ and $j+k-1$, i.e. operating in local product (physical) space
$\mathfrak{h}_{j}\otimes \cdots \mathfrak{h}_{j+k-1}$.
The steady state expectation values can then be expressed as
\begin{equation}
\expect{X_{j}}:=\frac{\Tr_{s}(X_{j}S_{n}S_{n}^{\dagger})}{\Tr_{s}(S_{n}S_{n}^{\dagger})}=
(\mathcal{Z}^{(n)}_{N})^{-1}\bra{0,0}(\mbb{T}_{N})^{j-1}\mbb{X}_{N}(\mbb{T}_{N})^{n-j-k+1}\ket{0,0},
\end{equation}
where the partial trace $\Tr_{s}$ is over $\mathfrak{H}_{s}$, and \textit{transition vertex operator} $\mbb{T}_{N}$ and $n$-particle \textit{non-equilibrium partition function}
\begin{equation}
\mathcal{Z}^{(n)}_{N}:=\bra{0,0}(\mbb{T}_{N})^{n}\ket{0,0}
\end{equation}
have been introduced.

\paragraph{Vertex operators.}
Particle density profiles are calculated by means of on-site elementary vertex operators $\{\mbb{V}_{N}^{(k)}\}$, $k=1,2,\ldots,N$, defined as,
\begin{equation}
\mbb{V}_{N}^{(k)}=\sum_{j=1}^{N}L^{jk}\otimes \ol{L}^{jk},
\label{density_vertex}
\end{equation}
enabling a compact definition of the transition operator,
\begin{equation}
\mbb{T}_{N}=\sum_{k=1}^{N}\mbb{V}_{N}^{(k)}.
\end{equation}
For brevity we shall drop the index $N$ from here on.

To each pair of local density operators $E^{kk}_{j}$ we may associate a current density operator $j^{(k,l)}$, defined in terms of a
\textit{local continuity equation} at site $j$,
\begin{equation}
\frac{d}{dt}\expect{E^{kk}_{j}-E^{ll}_{j}}=\expect{j^{(k,l)}_{j-1,j}-j^{(k,l)}_{j,j+1}}=0,\qquad k,l=1,2,\ldots,N
\end{equation}
explicitly reading
\begin{equation}
j^{(k,l)}_{j,j+1}=\ii(E^{kl}_{j}E^{lk}_{j+1}-E^{lk}_{j}E^{kl}_{j+1}).
\end{equation}
Introducing a corresponding $2$-site vertex operator $\mbb{J}^{(k,l)}=\ii(\mbb{J}^{(k,l)}_{+}-\mbb{J}^{(k,l)}_{-})$ and taking into account
that current expectation value must not depend of position (by virtue of continuity equation), we arrive at
\begin{equation}
\expect{j^{(k,l)}}=(\mathcal{Z}^{(n)}_{N})^{-1}\bra{0,0}(\mbb{T}_{N})^{n-2}\mbb{J}_{N}^{(k,l)}\ket{0,0},
\label{current_double}
\end{equation}
with 
\begin{equation}
\mbb{J}^{(k,l)}=\ii \sum_{i,j=1}^{N}\left[L^{il}L^{jk}\otimes \ol{L}^{ik}\ol{L}^{jl}-
L^{ik}L^{jl}\otimes \ol{L}^{il}\ol{L}^{jk}\right].
% old - with the definition E^{ij}(x)L^{ij}
%\mbb{J}^{(k,l)}=\ii \sum_{i,j=1}^{N}\left[\bb{L}^{li}\bb{L}^{kj}\otimes \ol{\bb{L}}^{ki}\ol{\bb{L}}^{lj}-
%\bb{L}^{kj}\bb{L}^{li}\otimes \ol{\bb{L}}^{lj}\ol{\bb{L}}^{ki}\right].
\end{equation}
Note that products of elements $L^{ij}$ above cannot be simplified since they do not consist of spin generator in the fundamental
representation, for which \eref{fundamental_comm_relations} holds.

\subsection{Vertex operator algebra for spin-$1/2$ chain}
\paragraph{Spin current.}
In this section we state some explicit results for the simplest $N=2$ case.
It is helpful to notice that transition vertex $\mbb{T}$ preserves the subspace spanned by \textit{diagonal states}
$\ket{k,k}\equiv (x^{2}_{1})^{k}(y^{2}_{1})^{k}=:z^{k}$ ($k\in \ZZ_{+}$), which allows us to reduce the expression \eref{current_double} by virtue of
the reduced vertex (hatted) operators $\hat{\mbb{T}}$ and $\hat{\mbb{J}}$, obtained through orthogonal projections onto subspace
$\mathfrak{H}_{{\rm diag}}={\rm span}\{z^{k}\equiv \ket{k};z\in \ZZ_{+}\}$, 
\begin{equation}
\expect{j^{(k,l)}}=(\hat{\mathcal{Z}}^{(n)})^{-1}\bra{0}\hat{\mbb{T}}^{n-2}\hat{\mbb{J}}\ket{0}.
\end{equation}
Expressing how $\hat{\mbb{T}}$ and $\hat{\mbb{J}}$ operate in $\mathfrak{H}_{{\rm diag}}$,
\begin{eqnarray}
\fl \sinc{\gamma}^{-2}\hat{\mbb{T}}&=\left([k-s+\ii \lambda]_q[k-\ol{s}-\ii \ol{\lambda}]_q+[k-s-\ii \lambda]_q[k-\ol{s}+\ii \lambda]_q\right)\ket{k}\bra{k}\nonumber \\
&~~~~+[k+1]^{2}_{q}\ket{k}\bra{k+1}+[k-2s]_q[k-2\ol{s}]_q\ket{k+1}\bra{k},\\
\fl \sinc{\gamma}^{-4}\hat{\mbb{J}}&=\Big(\ii[k-s+\ii \lambda]_q[k-s-\ii \lambda]_q[k+1]_q[k-2\ol{s}]\nonumber \\
&~~~~+\ii[k-\ol{s}-\ii \ol{\lambda}]_q[k-\ol{s}+\ii \ol{\lambda}]_q[k]_q[k-1-2s]_q\nonumber \\
&~~~~-\ii[k-\ol{s}-\ii \ol{\lambda}]_q[k-\ol{s}+\ii \ol{\lambda}]_q[k+1]_q[k-2s]_q\nonumber \\
&~~~~-\ii[k-s+\ii \lambda]_q[k-s-\ii \lambda]_q[k]_q[k-1-2s]_q\Big)\ket{k}\bra{k}\nonumber \\
&~~~~+\ii[k+1]^{2}_{q}\Big([k-s+\ii \lambda]_q[k+1-\ol{s}-\ii \ol{\lambda}]_q\nonumber \\
&~~~~-[k+1-s+\ii \lambda]_q[k-\ol{s}-\ii \ol{\lambda}]_q\Big)\ket{k}\bra{k+1}\nonumber \\
&~~~~+\ii[k-2s]_q[k-2\ol{s}]_q\Big([k-s-\ii \lambda]_q[k+1-\ol{s}+\ii \ol{\lambda}]_q\nonumber \\
&~~~~-[k+1-s-\ii \lambda]_q[k-\ol{s}+\ii \ol{\lambda}]_q\Big)\ket{k+1}\bra{k},
% with old definition a-la Derkachov
% \fl \sinc{\gamma}^{-2}\mbb{T}&=\left([k+s+\ii \lambda]_q[k+\ol{s}-\ii \ol{\lambda}]_q+[k+s-\ii \lambda]_q[k+\ol{s}+\ii \lambda]_q\right)\ket{k}\bra{k}\nonumber \\
% &~~~~+[k+1]^{2}_{q}\ket{k}\bra{k+1}+[k+2s]_q[k+2\ol{s}]_q\ket{k+1}\bra{k},\\
% \fl \sinc{\gamma}^{-4}\mbb{J}&=\Big(\ii[k+s+\ii \lambda]_q[k+s-\ii \lambda]_q[k+1]_q[k+2\ol{s}]\nonumber \\
% &~~~~+\ii[k+\ol{s}-\ii \ol{\lambda}]_q[k+\ol{s}+\ii \ol{\lambda}]_q[k]_q[k-1+2s]_q\nonumber \\
% &~~~~-\ii[k+\ol{s}-\ii \ol{\lambda}]_q[k+\ol{s}+\ii \ol{\lambda}]_q[k+1]_q[k+2s]_q\nonumber \\
% &~~~~-\ii[k+s+\ii \lambda]_q[k+s-\ii \lambda]_q[k]_q[k-1+2s]_q\Big)\ket{k}\bra{k}\nonumber \\
% &~~~~+\ii[k+1]^{2}_{q}\Big([k+s+\ii \lambda]_q[k+1+\ol{s}-\ii \ol{\lambda}]_q\nonumber \\
% &~~~~-[k+1+s+\ii \lambda]_q[k+\ol{s}-\ii \ol{\lambda}]_q\Big)\ket{k}\bra{k+1}\nonumber \\
% &~~~~+\ii[k+2s]_q[k+2\ol{s}]_q\Big([k+s-\ii \lambda]_q[k+1+\ol{s}+\ii \ol{\lambda}]_q\nonumber \\
% &~~~~-[k+1+s-\ii \lambda]_q[k+\ol{s}+\ii \ol{\lambda}]_q\Big)\ket{k+1}\bra{k},
\end{eqnarray}
one may observe that $\hat{\mbb{T}}$ and $\hat{\mbb{J}}$ are in fact \textit{proportional} to each other provided $\Re(\lambda)=0$
($s=\Re(s)+\ii\,{\rm Im}(s)$ can still be arbitrary though),
\begin{equation}
\label{J_propto_T}
\hat{\mbb{J}}=\rho_{\gamma}({\rm Im}(s))\, \hat{\mbb{T}},\quad
%\rho_{\gamma}(s):=-\sinc{\gamma}^{2}\frac{\sinh{(2\gamma s_{i})}}{\sin{\gamma}},
\rho_{\gamma}(s):=\ii\sinc{\gamma}^{2}[2\ii\, {\rm Im}(s)]_q.
\end{equation}
With aid of this result we find (as noted in \cite{PRL107}) for the expectation value of the (spin) current density,
\begin{equation}
\expect{j^{(k,l)}}=\rho_{\gamma}({\rm Im}(s))\frac{\hat{\mathcal{Z}}^{(n-1)}}{\hat{\mathcal{Z}}^{(n)}},
\label{current_ratio}
\end{equation}
which interestingly coincides with analogous expression known from exact solutions of classical ASEP~\cite{Derrida},
despite the latter are governed by so-called \textit{reaction-diffusion} (quadratic) algebras~\cite{Alcaraz93,Aneva}.
Essentially, the equation \eref{current_ratio} is implied by the local continuity equation and a \textit{weaker} requirement, namely
\begin{equation}
\label{T_propto_J_boundary}
\hat{\mbb{T}}\ket{0}\sim\hat{\mbb{V}}\ket{0},\qquad \bra{0}\hat{\mbb{T}}\sim\bra{0}\hat{\mbb{V}}
\end{equation}
%Consequently, the \textit{trace-preservation} condition, ${\rm Im}(\lambda)={\rm Im}(2r_{0})+{\rm Im}(r_{1})=0$,
%together with \textit{sufficient} condition for the validity of the continuity equation, finally imply that the spectral parameter $\lambda$ must indeed be set to zero.
which is important in the $SU(N)$ case where the proportionality relation \eref{J_propto_T} breaks down while
the partial contractions \eref{T_propto_J_boundary} remain valid, with the proportionality factor of $-\frac{8}{(N-1)^{2}\Gamma}$.

For non-generic, i.e. reducible cases, with deformation parameter $\gamma=\pi(l/m)$, one could in principle access a dense set of
anisotropy parameters in the easy-plane regime $|\cos{(\gamma)}|<1$ by means of diagonalization of the reduced transition operator $\hat{\mbb{T}}$, acting
in $m$-dimensional invariant subspace in $\mathfrak{H}_{{\rm diag}}$ spanned by states $\{\ket{k,k};k=0,\ldots,m-1\}$, as long as $m$ is
sufficiently small. In this sense, the asymptotic of \eref{current_ratio} is dominated by the largest eigenvalue.
For the undeformed (critical) $q=1$ case however, $\hat{\mbb{T}}$ is irreducible infinitely-dimensional operator, and thus an eigenproblem for $\hat{\mbb{T}}$ seems to
require analytic treatment in the spirit of approaches employed in studies of algebras associated to ASEP~\cite{Blythe,Aneva}.\footnote{There is
nonetheless one apparent difference in compare to ASEP, namely in our case the transition operator takes place in
the product representation of quantum algebra.}
We shall abstain from this technical aspect at the moment, going beyond the scope of this paper, thus leave it open for future analysis.

\paragraph{Particle density profiles.}
Computation of particle density profiles in thermodynamic ($n\to \infty$) limit can be on the other hand assisted with help of
closed algebraic relations among vertex operators $\hat{\mbb{T}}$ and $\{\hat{\mbb{V}}^{(k)}\}$, which we have found with assistance of symbolic
algebra. In particular, by defining the magnetization density vertex operator $\hat{\mbb{V}}^{(z)}:=\hat{\mbb{V}}^{0}-\hat{\mbb{V}}^{(1)}$,
the following \textit{third-order} relations from the free algebra of vertex operators $\{\hat{\mbb{T}},\hat{\mbb{V}^{(z)}}\}$ can be found:
\begin{equation}
[\hat{\mbb{T}},[\hat{\mbb{T}},\hat{\mbb{V}}^{(z)}]]=\kappa^{0}_{\gamma}(s_{i})\hat{\mbb{V}}^{(z)}+\kappa^{1}_{\gamma}(s_{i})\{\hat{\mbb{T}},\hat{\mbb{V}}^{(z)}\},
\end{equation}
for $\lambda=0$ and $s=\ii\,{\rm Im}(s)$, with coefficient functions
\begin{eqnarray}
\kappa^{0}_{\gamma}({\rm Im}(s))&=8\sinc{\gamma}^{4}[\ii\, {\rm Im}(s)]^{2}_{q}\cos{(2\gamma)}\cosh{(2\gamma {\rm Im}(s))},\nonumber \\
\kappa^{1}_{\gamma}({\rm Im}(s))&=-2\sinc{\gamma}^{2}\cos{(2\gamma)},
\end{eqnarray}
reducing in the undeformed limit $\gamma \rightarrow 0$ to $\kappa^{0}_{0}(s_{i})=-8\,({\rm Im}(s))^{2}$, $\kappa^{1}_{0}=-2$, as stated
previously in \cite{PRL107}.

It remains another appealing problem how to understand those type of relations (and eventually their counterparts for higher $N$)
from the algebraic standpoint, whereas from practical perspective, to provide a method for evaluating particle density profiles for finite
chains of size $n$ and/or obtaining closed-form results in the thermodynamic regime from first symmetry principles
(i.e. without resorting on model-specific asymptotic observation~\cite{PRL107}).

\paragraph{Density profiles.}
By means of explicit symbolic contraction for finite system sizes $n$ (where effective size of the auxiliary space $\mathfrak{H}_{s}$ becomes
finite) we have computed the particle density profiles for each particle species for $N$-dimensional case, and \textit{conjectured}
the following \textit{asymptotic} form,
\begin{eqnarray}
\expect{E^{kk}_{j}}=\frac{1}{2(N-1)}\left[1+\cos{\left(\frac{j\pi}{n}\right)}\right],\qquad k=1,2,\ldots,N-1,\nonumber \\
\expect{E^{NN}_{j}}=-(N-1)\expect{E^{kk}_{j}},\qquad k\neq N
% finite size N-dependent correction
%\expect{n^{k}_{j}}=\frac{1}{2(N-1)}\left[1+\cos{\left(\frac{j\pi}{n-(N-1)^{-1}}\right)}\right],
\end{eqnarray}
which for $N=2$, i.e. spin-$1/2$ case, correctly reduces to previously found cosine-shaped magnetization profiles~\cite{PRL107}.

\section{Conclusions and discussion}

We provided a unification of exact MPO solutions of nonequilibrium steady states for (boundary-driven)
dissipative quantum evolution with integrable bulk Hamiltonians, just recently appearing in the literature. We explicitly displayed how
the solutions originate from rudimentary concepts of quantum theory of integrability, namely the Yang-Baxter equation and representation theory of quantum groups, ensuring a suitable cancellation mechanism (known as the Sutherland equation) for the action of
the unitary part of Liouville super-operator. In the second stage, the remaining surface-like terms enter into the system of boundary compatibility
conditions which are subsequently treated independently of the bulk structure in combination with general form of the dissipation rate-matrix,
and with suitably chosen boundary states.
Although we were able to find a family of solutions belonging to $SU(N)$-invariant quantum gases, by limiting ourselves to a restricted set of
dissipative channels and adopting boundary vacuum states, it still remains an open problem to systematically classify all solutions for a given bulk interaction.

Moreover, we asserted how lowest weight transfer matrices, defined in terms of non-unitary $\mathfrak{sl}(N,\CC)$ realizations of ancillary
spaces, in a sense provide a natural setup for a description of nonequilibrium states of open systems in absence translational invariance.
As it has been recently pointed out~\cite{PIP}, presented exact solutions offer an interesting perspective into theory of quantum transport~\cite{Zotos97}.
In the first place, they open a possibility for analytic studies of paramount physical properties (e.g. transport coefficients, phase transitions, fluctuation theorems)
of quantum out-of-equilibrium processes and their steady states by means of well-understood underlying symmetry principles,
along the lines of their classical counterparts, to obtain some fresh closed-form results. Quite remarkably, in spite of intrinsic
far-from-equilibrium character of the problem, solutions at hand also provide a valuable insight on transport theory within linear-response regime,
as they essentially (by construction) represent a continuous family of \textit{quasi-local conserved charges}~\cite{QLAC}, complementing
an infinite sequence of \textit{local} (standard) conservation laws, having a profound influence on anomalous transport behavior
in quantum spin chains~\cite{PRL106,Drude}. Quite strikingly, most recent numerical evidence~\cite{macroscopic} reveals that quantum integrability structure,
giving birth to quasi-local integrals of motion most likely survives a passage to the \textit{classical} integrable continuous models and corresponding
integrable lattice regularizations, indicating that part of presented integrability structures indeed have well-defined classical counterparts.

Finally, we give some remarks on limitations of the presented framework, apart from inability of extending the formalism to accommodate for the
non-fundamental models, as we have already argued. Requiring solutions exhibiting an underlying
quantum group symmetry seems to be inessential from purely algebraic requirements imposed by the bulk and the boundary conditions
for steady NESS density operators, since merely a \textit{weaker} condition imposed by the Sutherland-type of equation is in fact sufficient
in the bulk. In this regard, boundary matrix need not necessarily be related to the derivative of the Lax matrix with respect to the spectral parameter. On the other hand, we
could of course seek for algebraic constructions not strictly of quantum group type, e.g. some other quadratic algebras admitting infinite-dimensional realizations.
That being said, one could address elliptic ($8$-vertex) solutions of the YBE, with no associated quasi-triangular Hopf algebra,
\footnote{Instead, one can construct algebraic objects called elliptic quantum groups~\cite{Felder}, which are two-parametric deformations of
$\mathcal{U}(\mathfrak{gl}(N,\CC))$.}
describing a local Hamiltonian without particle conservation law (e.g. $XYZ$ Heisenberg model), implemented with aid of a Lax matrix formed
from variables constituting the Sklyanin's quadratic algebra~\cite{Sklyanin82}, curiously also permitting to construct infinite-dimensional
'continuous-spin' representations~\cite{Zabrodin}, in analogy to Verma modules presented in \Sref{sec:Verma}.

Beside that, a constrained Cholesky-type factorized form of the density operator in terms of an abstract (non-hermitian) transfer matrix,
does presumably not exhaust all possibilities within boundary-driven setup. Eventually, we could have included coherent ultra-local boundary
fields~\cite{KPS}, which might turn out to have an important role in bulk-boundary compatibility condition.
Exploring these directions could be an interesting option for further developments.

\clearpage
\renewcommand{\theequation}{\thesection.\arabic{equation}}
\begin{appendices}

\section{Twisting solutions}
\label{App:twist}
We show how non-hermitian terms in the Hamiltonian density can be removed by twisting the Hopf algebra structure,
formally expressed as a similarity transformation of the coproduct while preserving coassociativity property,
\begin{equation}
\Delta(\xi)\rightarrow \Delta_{t}(\xi)=\mathcal{F}_{12}\Delta(\xi)\mathcal{F}_{12}^{-1},\quad
\mathcal{F}_{12}\mathcal{F}_{13}\mathcal{F}_{23}=\mathcal{F}_{23}\mathcal{F}_{13}\mathcal{F}_{12}.
\end{equation}
The universal $\mathcal{R}$-matrix then transforms as
\begin{equation}
\mathcal{R}\rightarrow \mathcal{R}_{t}=\mathcal{F}_{12}^{-1}\mathcal{R}\mathcal{F}_{12}^{-1}.
\end{equation}
Recall how non-hermicity arose as a consequence of Baxterization of the parameter-less universal $\mathcal{R}$-matrix, with the introduction
of the $\lambda$-dependent factors in its off-diagonal elements, producing extra unwanted terms in the interaction
$h\sim \partial_{\lambda}\PR(\lambda)|_{\lambda=0}$. Following the procedure proposed in \cite{Bytsko}, the fundamental $\Uqsl{2,\CC}$
$\mathcal{R}$-matrix transforms via $\beta$-dependent diagonal universal twisting
element $\mathcal{K}_{\lambda}=\exp{(\beta \lambda S^{z})}$ in the space $\mathfrak{S}_{s}$
\begin{equation}
\mathcal{R}_{\beta}(\lambda)=(\mathds{1}\otimes \mathcal{K}_{\beta}(\lambda))\mathcal{R}(\lambda)(\mathds{1}\otimes \mathcal{K}_{\beta}(\lambda))^{-1}.
\label{universal_twist}
\end{equation}
Evaluating \eref{universal_twist} in the fundamental representation $\mathfrak{S}_{f}$,
i.e. using $K^{(f)}_{\beta}(\lambda)=\exp{(\gamma \lambda/2)\sigma^{z}}$,
with the choice $\beta=\gamma$, yields
\begin{eqnarray}
R^{(f,f)}_{\gamma}(\lambda)&=(\mathds{1}\otimes K^{(f)}_{\lambda})R^{(f,f)}(\lambda)(\mathds{1}\otimes K^{(f)}_{\lambda})^{-1}\nonumber \\
&=(q-q^{-1})\pmatrix{
[-\ii \lambda+1]_q & & & \cr
& [-\ii \lambda]_q & & \cr
& & [-\ii \lambda]_q & \cr
& & & [-\ii \lambda+1]_q
}.
\end{eqnarray}
Accordingly, the $L$-operator can modified by twisting in the quantum space $\mathfrak{S}_{f}$ only,
\begin{equation}
\bb{L}_{\beta}(\lambda)=e^{\beta \lambda S^{z}}\bb{L}(\lambda)e^{-\beta \lambda S^{z}},
\end{equation}
being an automorphism of $\Uqsl{2,\CC}$ algebra, $S^{\pm}\rightarrow e^{\pm \beta \lambda}S^{\pm},S^{z}\rightarrow S^{z}$,
yielding at $\beta=\gamma$ the standard $\Uqsl{2,\CC}$-covariant $L$-operator of the $XXZ$ Heisenberg model
\begin{equation}
\bb{L}^{XXZ}(\lambda)=\pmatrix{
[-\ii \lambda + S^{z}]_q & S^{-} \cr
S^{+} & [-\ii \lambda - S^{z}]_q}.
\end{equation}
It is noteworthy that applied $\beta$-twist breaks the $\Uqsl{2,\CC}$ symmetry of the (non-hermitian) interaction \eref{nonhermitian_interaction}
down to $U(1)$.

\subsection{Twisted Heisenberg model: $\Theta$-XXZ model}
Twist transformations allow to introduce extra parameter dependence into solutions of the YBE (see e.g. \cite{KunduRev}).
As a simple example we show how to transform the trigonometric $6$-vertex $R$-matrix by means of a diagonal matrix depending
on angle parameter $\Theta\in[0,2\pi)$.
In order to generate the \textit{asymmetric} (Wu-McCoy) anisotropic Heisenberg model with
\textit{vector-like} (Dzyaloshinkii-Moriya) interaction -- we have to choose an \textit{abelian} (Reshetikhin) twist \cite{Reshetikhin90,Dimitrijevic09},
\begin{equation}
\mathcal{F}_{\Theta}=e^{-(\ii \Theta/2)(S^{z}\otimes \mathds{1}-\mathds{1}\otimes S^{z})},
\end{equation}
evaluated in the product of fundamental representations $\mathfrak{S}_{f}\otimes \mathfrak{S}_{f}$,
\begin{equation}
F^{(f,f)}_{\Theta}={\rm diag}(1,e^{-\ii \Theta/2},e^{\ii \Theta/2},1),
\end{equation}
producing the following colored trigonometric $6$-vertex $R$-matrix ($\varphi:=-\ii \gamma \lambda$)
\begin{eqnarray}
\PR^{(f,f)}_{\Theta}(\lambda)=P_{12}R^{(f,f)}_{\Theta}(\lambda)=P_{12}F^{(f,f)}_{\Theta}R^{(f,f)}(\lambda)F^{(f,f)}_{\Theta}=\\
\frac{2\ii}{\gamma}\pmatrix{
\sin{(\varphi+\gamma)} & 0 & 0 & 0 \cr
0 & \sin{\gamma} & e^{\ii \Theta}\sin{\varphi} & 0 \cr
0 & e^{-\ii \Theta}\sin{\varphi} & \sin{\gamma} & 0 \cr
0 & 0 & 0 & \sin{(\varphi+\gamma)}}.
\end{eqnarray}
Taking derivative with respect to $\lambda$ at $\lambda=0$, we extract the interaction with $\Theta$-modified hopping term,
\begin{equation}
h^{\Theta}_{12}=2(e^{\ii \Theta}\sigma^{+}_{1}\sigma^{-}_{2}+e^{-\ii \Theta}\sigma^{-}_{1}\sigma^{+}_{2})+
2\cos{(\gamma)}\sigma^{z}_{1}\sigma^{z}_{2}+const.,
\label{Theta-XXZ}
\end{equation}
whereas the Lax operator gets transformed into its $\Theta$-twisted correspondent via universal twist $\mathcal{F}_{\Theta}$ evaluated
in $\mathfrak{S}_{f}\otimes \mathfrak{S}_{s}$, i.e.
$\bb{F}^{(f,s)}_{\Theta}=e^{-\ii (\Theta/4)(\sigma^{z}\otimes \mathds{1})}e^{-\ii (\Theta/2)(\sigma^{0}\otimes S^{z})}$,
\begin{eqnarray}
\bb{L}_{\Theta}(\lambda)&=\bb{F}^{(f,s)}_{\Theta}\bb{L}(\lambda)\bb{F}^{(f,s)}_{\Theta}= \\
&=\sinc{\gamma}
\pmatrix{
[-\ii \lambda+S^{z})]_q\,e^{\ii \Theta (S^{z}-\half)} & e^{\ii \Theta (S^{z}+\half)}S^{-} \cr
S^{+}e^{\ii \Theta (S^{z}+\half)} & [-\ii \lambda-S^{z})]_q\,e^{\ii \Theta (S^{z}+\half)}
}.\nonumber
\end{eqnarray}

The generalized anisotropic Hamiltonian \eref{Theta-XXZ} which introduces an electric-field term (meanwhile reducing hopping interaction)
brings a possibility of studying an interplay between spin currents contributed from the bulk interaction and external magnetization bias,
avoiding using twisted boundary fields~\cite{Alcaraz88,Brockmann13}. Curiously, regardless of parameter $\Theta$, the solution to the boundary equations
\eref{boundary_compact} remains unaffected, therefore defining a continuous family of solutions for spin-$1/2$ chain with maximal
incoherent driving dissipation provided in \Sref{sec:solutions}.

% interesting Theta does not affect boundary equations, any hence defines a continous family of solution generalizing maximally-driven
% s=1/2 solution

% REMARKS: 
% U(1) symmetry for R-matrix of XXZ model [R_{a_{1},a_{2}},S^{z}_{tot}]=0, comment also how interaction's symmetry is reduced
% from Uq(sl(2)) to U(1)

\section{Commuting transfer matrices}
\label{app:commuting}

As we have already argued in \Sref{sec:solutions}, the introduced $S$-operators factorizing NESS density operators $\rho_{\infty}$,
are in fact, unlike operators $\rho_{\infty}$ themselves, abstract \textit{non-hermitian} quantum transfer matrices subjected to
open boundary conditions. This property is attributed to existence of the universal $R$-matrix for the Lax operator pertaining to
general \textit{non-compact} spin representations. We shall briefly outline their origin in this section.
In this regard, we might (loosely speaking) proclaim the type of solutions presented in the preceding discussion as
\textit{integrable steady states}, based on the fact that they can be thought of as two fused (lowest weight) transfer matrices.
%\footnote{More information-theoretic inspired depiction would be to understand NESS density operator as a contracted ladder tensor
%network with two integrable legs.}
Pretty remarkably, those infinite-dimensional spin representations have previously found their places and applications in
supersymmetric quantum field theories, in particular in the integrable sector of the high-energy QCD~\cite{Korchemsky}.
Here, on contrary, we emphasize their importance in paradigmatic models of (non-relativistic) quantum statistical mechanics, i.e. in the
realm of non-canonical mixed-states associated to driven open (one-dimensional) quantum systems.

\subsection{Universal $R$-matrix for arbitrary complex-valued spin}
We begin by acknowledging a general type of solution of YBE over triple-product space
$\mathfrak{S}_{s_{1}}\otimes \mathfrak{S}_{s_{2}}\otimes \mathfrak{S}_{s_{3}}$,
\begin{equation}
R_{12}^{(s_{1},s_{2})}(\lambda-\mu)R_{13}^{(s_{1},s_{3})}(\lambda)R_{23}^{(s_{2},s_{3})}(\mu)=
R_{23}^{(s_{2},s_{3})}(\mu)R_{13}^{(s_{1},s_{3})}(\lambda)R_{12}^{(s_{1},s_{2})}(\lambda-\mu),
\label{general_YBE}
\end{equation}
investigated in \cite{arbitrary,Faddeev1,Faddeev2,integral,KKM}. We shall refrain from using our boldface convention henceforth.
For instance, setting \textit{all three} representation parameter to $s_{j}=\half=f$,
i.e. evaluating \eref{general_YBE} in the three-fold fundamental space $\mathfrak{S}^{\otimes 3}_{f}$, yields the fundamental
Yang-Baxter equation
\begin{equation}
R^{(f,f)}_{12}(\lambda-\mu)R^{(f,f)}_{13}(\lambda)R^{(f,f)}_{23}(\mu)=R^{(f,f)}_{23}(\mu)R^{(f,f)}_{13}R^{(f,f)}_{12}(\lambda-\mu),
\end{equation}
with well-known $\mathfrak{sl}(2,\CC)$-invariant solution $R^{(f,f)}_{12}(u)=u+P_{12}$, with permutation operator $P_{12}$ over product of
fundamental spaces $\mathfrak{S}_{f}\otimes \mathfrak{S}_{f}\cong \CC^{2}\otimes \CC^{2}$. Next, choosing $s_{1}=s_{2}=f$ and $s_{3}=s$,
we have, beside the standard rational $4\times 4$ $R$-matrix $R^{(f,f)}(u)$, another operator $R^{(f,s)}(u)$, acting in
$\mathfrak{S}_{f}\otimes \mathfrak{S}_{s}$, explicitly reading
\begin{equation}
R^{(f,s)}(u)=\left(u+\half\right)\mathds{1}+\vec{\sigma} \otimes \vec{S}=
\pmatrix{
(u+\half)\mathds{1}+S^{z} & S^{-} \cr
S^{+} & (u+\half)\mathds{1}+S^{z}
}.
\end{equation}
The latter is essentially nothing but a standard Lax operator for spin-$s$ (with $\mathfrak{S}_{f}\cong \CC^{2}$ auxiliary space)
from the $RLL$-relation, modulo additive constant and shift in the spectral parameter,
\begin{equation}
L^{(f,s)}(u)=R^{(f,s)}\left(u-\half \right),
\end{equation}
being equivalent to defining relations of spin-$s$ generators.
The third form is obtained by realizing only one space in the fundamental representation, say $s_{3}=f$. In this case we find YBE in the form
\begin{equation}
R^{(s_{1},s_{2})}_{12}(\lambda-\mu)R^{(s_{1},f)}_{13}(\lambda)R^{(s_{2},f)}_{23}(\mu)=
R^{(s_{2},f)}_{23}(\mu)R^{(s_{1},f)}_{13}R^{(s_{1},s_{2})}_{12}(\lambda-\mu),
\label{ourYBE}
\end{equation}
which involves a general $\mathfrak{sl}(2,\CC)$-invariant $R$-operator,
\begin{equation}
[\vec{S},R^{(s_{1},s_{2})}(u)]=0,
\label{sl2_invariance}
\end{equation}
over product space $\mathfrak{S}_{s_1}\otimes \mathfrak{S}_{s_2}$ of two arbitrary representations.
Those general solutions, addressed from representation-theoretic point of view in references \cite{Tarasov83,KRS}, ensure mutual commutativity
for general-type transfer matrices, obtained as (regularized) traces over generic spin representations \cite{Bazhanov10}, and are thus related to
non-compact spin chains~\cite{Korchemsky,Beisert03}. In parallel, solutions to \eref{ourYBE} have been constructed for lowest
weight representations -- assuming $s_{1},s_{2}\not \in \half \mathbb{Z}_{+}$, where finite-dimensional representation
theory applies~\cite{KRS,Faddeev1,Faddeev2} -- by means of eigenspace decomposition
\begin{equation}
R^{(s_{1},s_{2})}(u)\sim \sum_{\nu=0}^{\infty}r_{\nu}(u)P_{\nu},\qquad
R^{(s_{1},s_{2})}(u)\psi^{0}_{\nu}=r_{\nu}(u)\psi^{0}_{\nu},
\end{equation}
where $P_{\nu}$ is a projector from $\mathfrak{S}_{s_{1}}\otimes \mathfrak{S}_{s_{2}}$ to subspace $\mathfrak{S}_{s_{1}+s_{2}+\nu}$,
via recurrence relation
\begin{equation}
% a-la Dobrev
(u+s_{1}+s_{2}-\nu)r_{\nu+1}(u)=-(-u+s_{1}+s_{2}-\nu)r_{\nu}(u),
% Derkachov
%(-u+s_{1}+s_{2}+\nu)R_{\nu+1}(u)=-(u+s_{1}+s_{2}+\nu)R_{\nu}(u),
\end{equation}
with explicit solution,
\begin{equation}
% a-la Dobrev
r_{\nu}(u)=(-1)^{\nu}\frac{\Gamma(u+s_{1}+s_{2})\Gamma(-u+s_{1}+s_{2}-\nu)}{\Gamma(-u+s_{1}+s_{2})\Gamma(u+s_{1}+s_{2}-\nu)},
% Derkachov
%R_{\nu}(u)=(-1)^{\nu}\frac{\Gamma(-u+s_{1}+s_{2})\Gamma(u+s_{1}+s_{2}+\nu)}{\Gamma(u+s_{1}+s_{2})\Gamma(-u+s_{1}+s_{2}+\nu)},
\end{equation}
using normalization such that $R^{(s_{1},s_{2})}(u)\ket{0}\otimes \ket{0}=\ket{0}\otimes \ket{0}$, where $\ket{0}\otimes \ket{0}=\psi^{0}_{0}=1$.

On the other hand, in absence of cyclic invariance, our $S$-operator is defined as a \textit{lowest weight contraction} of the
monodromy matrix $\bb{T}(\lambda)$ rather than a partial trace $\Tr_{a}$ over $\mathfrak{H}_{a}\cong \mathfrak{S}_{s_{3}}$,
also displays transfer matrix property. This intriguing result, being recently observed~\cite{PIP}
\footnote{In the paper authors utilize Lax matrix with manifestly broken $\mathfrak{sl}(2,\CC)$ symmetry and construct corresponding
\textit{non-universal} $R$-matrix relying of somehow weaker symmetry principles.}, is based on preservation of left/right lowest weight
states $\bra{0,0}\equiv \bra{0}\otimes \bra{0}$ and $\ket{0}\equiv \ket{0}\otimes \ket{0}$,
\begin{equation}
\PR^{(s_{1},s_{2})}(u)\ket{0,0}=\ket{0,0},\qquad \bra{0,0}\PR^{(s_{1},s_{2})}(u)=\bra{0,0},
\end{equation}
implying
\begin{eqnarray}
S_{n}(\lambda,s_{1})S_{n}(\mu,s_{2})&=\bra{0,0}\bb{T}^{(s_{1})}_{1}(\lambda)\bb{T}^{(s_{2})}_{2}(\mu)\ket{0,0}=\nonumber \\
&=\bra{0,0}\PR^{(s_{1},s_{2})}_{12}(\lambda-\mu)\bb{T}^{(s_{1})}_{1}(\lambda)\bb{T}^{s_{2}}_{2}(\mu)\ket{0,0}=\nonumber \\
&=\bra{0,0}\bb{T}^{(s_{2})}_{1}(\mu)\bb{T}^{(s_{1})}_{2}(\lambda)\PR^{(s_{1},s_{2})}_{12}(\lambda-\mu)\ket{0,0}=\nonumber \\
&=S_{n}(\mu,s_{2})S_{n}(\lambda,s_{1}).
\label{commutation_property}
\end{eqnarray}

Although we focused entirely on non-deformed case, above consideration is applicable to $q$-analogues as well~\cite{KKM}, and in addition
to the Lie algebra $\mathfrak{gl}(N,\CC)$ with $N$ complex parameters (or even superalgebra),
namely a central element associated to the spectral parameter and the remaining $(N-1)$ $\mathfrak{sl}(N,\CC)$ representation parameters.

\subsection{Coherent boundary vectors}
Clearly, an extension to generalized transfer matrices contracted with left and right \textit{coherent states}
\begin{equation}
\ket{\phi_{R}(s)}:=e^{\phi_{R}S^{+}}\ket{0},\qquad \bra{\phi_{L}}:=\bra{0}e^{\phi_{L}S^{-}(s)},\qquad \phi_{L,R}\in \CC,
\end{equation}
is straightforward, i.e. a family of the $S$-operators
\begin{equation}
S_{n}(\lambda,s|\phi_{L},\phi_{R})\sim \bra{\phi_{L}(s)}\bb{T}^{(s)}(\lambda)\ket{\phi_{R}(s)},\qquad
\end{equation}
maintains the commutativity property,
\begin{equation}
[S_{n}(\lambda,s_{1}|\phi_{L},\phi_{R}),S_{n}(\mu,s_{2}|\phi_{L},\phi_{R})]=0,
\label{coherent_commute}
\end{equation}
simply as a consequence of the $\mathfrak{sl}(2,\CC)$ invariance \eref{sl2_invariance},
hence the argument from \eref{commutation_property} still applies.

\end{appendices}

\ack
Authors are grateful to T. Prosen for inspiring discussions, and particularly for reviewing the manuscript.
E. I. thanks V. Popkov for communicating the property \eref{coherent_commute}.
B. {\v Z}. acknowledges the support of FONDECYT project N${}^\circ$3130495.

\pagebreak
\section*{References}
\bibliographystyle{ieeetr}
\bibliography{QGA_beta.bib}

\end{document}